# Cine MRI–Validated Biventricular Electromechanical Digital Twin Framework for Predicting Regional Endocardial Motion

Milad Hasani, Tomas Zaremba, Sam Riahi, and Alireza Rezania

***Abstract*— Personalized cardiovascular medicine increasingly relies on digital twin frameworks to translate clinical imaging into patient-specific functional insights. This work presents a biventricular electromechanical human heart model for predicting regional endocardial motion. The model integrates realistic 3D cardiac geometry, rule-based myocardial fiber orientation, reaction–diffusion electrophysiology, voltage-dependent active stress, closed-loop systemic and pulmonary hemodynamics, and two-way fluid–structure interaction with explicit 3D blood domains. Mechanical boundary conditions are also introduced to represent the influence of surrounding tissues on cardiac motion. The model was validated against Cine magnetic resonance imaging (MRI)-derived right ventricular motion, demonstrating consistent regional motion patterns between simulation and imaging results. The validated framework enables quantitative assessment of regional endocardial displacement and velocity, supporting the identification of mechanically favorable implantation regions for motion-driven intracardiac devices. The results further highlight that implantation-site selection should consider not only local motion amplitude, but also anatomical safety and electrophysiological suitability. Overall, the proposed Cine MRI–validated electromechanical digital twin framework provides a predictive platform for regional endocardial motion analysis and establishes a foundation for future patient-specific planning of self-powered intracardiac implants prior to clinical implementation.**

***Index Terms*— Cardiac electromechanical model, magnetic resonance imaging (MRI), cardiac digital twin framework, fluid–structure interaction, intracardiac energy harvesting.**

## I. Introduction

Personalized cardiovascular medicine is undergoing a paradigm shift, driven by the convergence of advanced clinical imaging and computational engineering to create patient-specific predictive models. Patient-specific computational cardiac models serve as the computational foundation for cardiac digital twins (CDTs), which have become increasingly important in modern cardiology. A CDT is an integrative, multiscale computational model that provides a virtual representation of an individual's cardiac structure and function, incorporating clinical data to enable personalized assessments of pathologies and interventions [1]–[4]. These digital twin frameworks serve as powerful tools for enhancing the understanding of cardiac physiology and pathology, and for optimizing therapeutic interventions before they are implemented clinically [4]–[6].

Historically, CDTs have been developed to simulate a wide array of cardiac phenomena. A substantial body of literature has focused on modeling cardiac electrophysiology (EP), a fundamental process that governs the heart's electrical activation and its propagation through the myocardium [7]–[11]. EP models often employ systems of partial differential equations (PDEs) coupled with detailed ionic models to predict the spatio-temporal evolution of action potentials [8], [12]. The formulation of these models has also been adapted to account for factors like anisotropic conductivity, Purkinje network effects, and fiber orientation, which significantly influence wave propagation patterns [3], [6], [13]. Rule-based methods have been used to assign fiber orientation based on histological observations to improve the realism of simulations [12], [14]. These studies have also been extended to incorporate myocardial mechanics, describing how muscle fiber orientation, tissue elasticity, and active stress generation drive ventricular contraction and deformation [4], [15]–[17].

More advanced models involve coupled lumped-parameter networks for simulating both blood circulations [18], [19]. This approach enables the derivation of critical hemodynamic outputs, including pressure-volume loops, blood flow rates, and ejection fraction. These integrated models have proven valuable for investigating device optimization, such as left ventricular assist devices, and for assessing global ventricular function under various loading conditions [4]. However, to accurately account for the effects of blood flow within the chambers, intracardiac flow and fluid–structure interaction (FSI) should be examined by incorporating three-dimensional blood meshes. Because 3D blood flow and FSI analysis lead to significant computational costs, several previous works [18], [19] introduced a cavity-volume constraint so that ventricular

This work was supported by a research grant from the Danish Cardiovascular Academy, which is funded by the Novo Nordisk Foundation under grant NNF20SA0067242, and the Danish Heart Foundation. DeiC National HPC under grant DeiC-AAU-N2-2025122 provided the required computational resources. (Corresponding author: Milad Hasani).

Milad Hasani and Alireza Rezania are with AAU Energy, Aalborg University, Aalborg, Denmark (e-mail: miha@energy.aau.dk)
Tomas Zaremba is with the Department of Clinical Medicine, Aalborg University, Aalborg, Denmark, and the Department of Cardiology, Aalborg University Hospital, Aalborg, Denmark.
Sam Riahi is with the Department of Cardiology, Aalborg University Hospital, Aalborg, Denmark.

pressure acts as a Lagrange multiplier applied as endocardial traction.

While electromechanical models in the literature have been used to predict a variety of outputs, translating raw imaging data into accurate predictions of physiological heart motion remains a challenge. The wall heart motion is traditionally investigated via motion sensors [20]–[22] and medical image processing [23], [24]. To address this gap and advance personalized cardiovascular care, this work develops and validates a high-fidelity biventricular electromechanical digital twin framework designed specifically for regional endocardial motion prediction. To account for the effect of blood flow on heart motion, this research implemented actual 3D blood meshes and a two-way FSI formulation. The model also includes innovative epicardial boundary conditions to simulate the stiffness and viscosity of neighboring tissues.

Crucially for clinical translation, the predicted regional motion is validated against Cine MRI-derived right ventricular motion from an independent healthy-subject dataset, demonstrating the framework's ability to reproduce physiologically consistent regional motion patterns. By coupling excitation, deformation, and flow, the framework enables quantitative assessment of regional wall deformation, which can be useful for identifying mechanically favorable implantation sites for intracardiac devices, evaluating medical implants, and providing a foundation for future patient-specific therapeutic planning. For instance, for a self-powered leadless pacemaker implanted over the right ventricle, the implantation site plays a critical role, since the amount of harvestable energy depends strongly on the magnitude of local cardiac motion at the implant location [20], [25], [26].

This article is organized into four main sections. Section II details the development of the electromechanical model, covering the 3D human heart geometry, electrophysiology, electromechanical activation, two-way FSI, and mechanical boundary conditions. Section III presents the results of the model and its validation against Cine MRI scans. Finally, Section IV offers the conclusion and clinical outlook for the research.

## II. DEVELOPMENT OF THE ELECTROMECHANICAL DIGITAL TWIN FRAMEWORK

The development of the biventricular electromechanical digital twin framework involves a multifaceted approach that integrates detailed anatomical geometry with advanced computational models for electrophysiology, mechanics, and hemodynamics. The foundation of the model is a realistic 3D human heart geometry, derived from actual anatomical data. Building upon this, the anisotropic contraction and electrical conduction of the myocardium are defined by prescribing the spatial helical arrangement of cardiac fibers. Electrophysiological activity is governed by a reaction-diffusion framework, while the link between electrical excitation and mechanical contraction is modeled by a voltage-dependent active stress term. The entire 3D heart model is then coupled with a 0D closed-loop hemodynamic model that represents the systemic and pulmonary circulations. Finally, a two-way FSI formulation links the blood flow in the ventricular chambers to the contracting myocardium, with mechanical boundary conditions applied to mimic the influence of neighboring tissues.

### A. Anatomically Realistic 3D Human Heart Geometry

While many models in the literature are based on idealized or simplified heart geometries, the use of actual anatomical shapes is critical for accurately capturing the complex, localized motion patterns of the heart. Simplified geometries often fail to represent important anatomical variations such as ventricular asymmetries, wall thickness differences, and irregular myocardial contours, all of which play a significant role in the mechanical behavior of the heart during the cardiac cycle. In this study, a realistic 3D heart geometry was employed to ensure that localized motions, including subtle regional deformations and torsional effects, could be accurately derived.

To construct anatomically accurate 3D solid models of the human heart, data from the biventricular dataset provided by the Cardiac Atlas Project [27] were used. This dataset consists of spatially distributed data points representing the geometry of three anatomical components: the Right Ventricle (RV), the Left Ventricle (LV), and the Myocardium (Myo), all derived from cardiac MRI of real human subjects. These data point groups are first integrated into a unified coordinate system. Interpolation techniques are then applied to reconstruct continuous surface representations from the discrete points. Following this, mesh generation and solidification procedures are performed to obtain closed volumetric geometries suitable for computational modeling and simulation.

Fig. 1 illustrates the reconstruction workflow. Individual point clouds representing RV, LV, and Myo (shown in blue, red, and green, respectively) are combined to form a composite 3D point cloud. This composite is then processed into a smooth solid model of the heart consisting of RV and LV chambers and myocardium, enabling further anatomical or functional analysis.

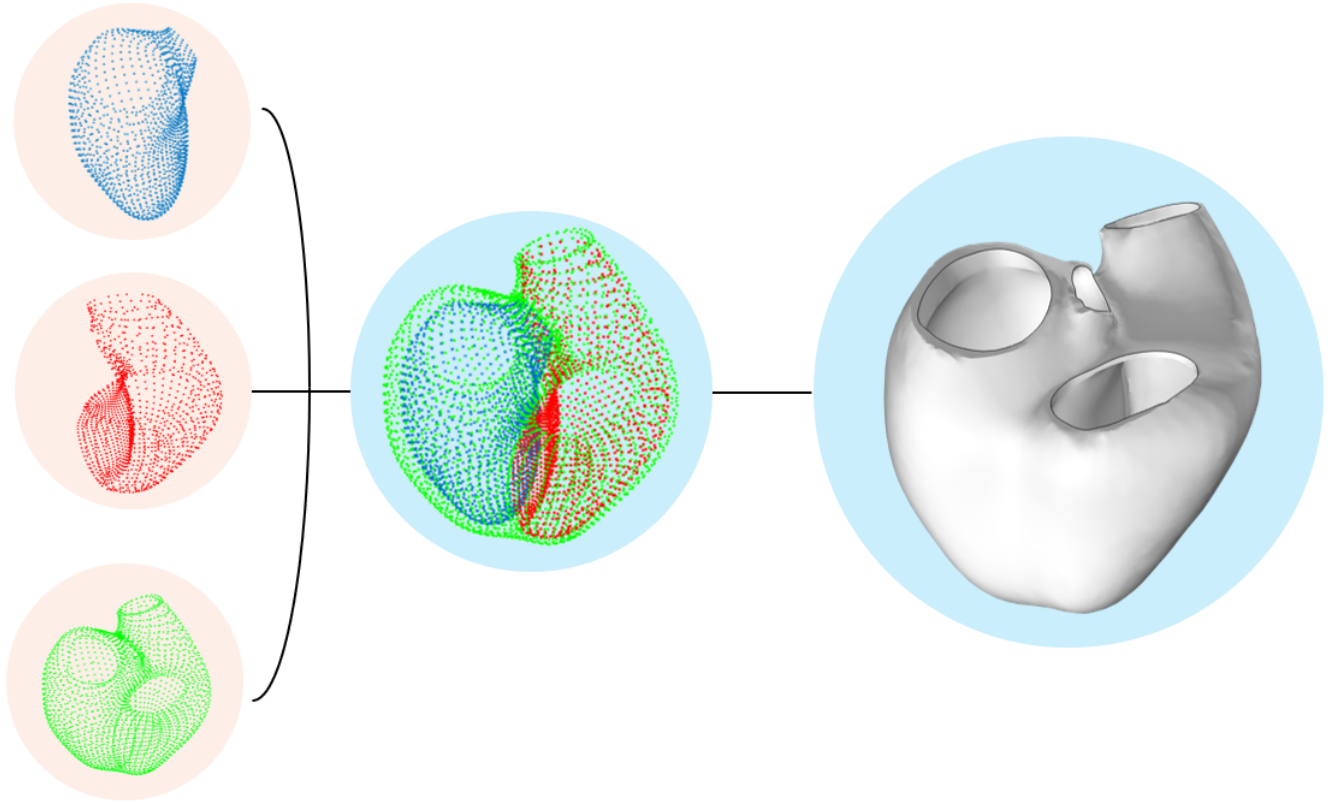

Fig. 1- Reconstruction of the 3D human heart geometry from the Cardiac Atlas Project dataset. Point clouds representing the LV (blue), RV (red), and Myocardium (green) were integrated and processed to produce a closed, anatomically realistic 3D heart geometry.

### B. Muscle Fibers Implantation

The anisotropic contraction and electrical conduction of the myocardium are governed by the spatial helical arrangement of cardiac fibers. The distribution of myofibers is highly complex and exhibits localized orientations that vary transmurally and regionally throughout the ventricular wall. The orientation of myofibers has been investigated through histological data [13], [28] and the diffusion tensor imaging method [29], [30]. Histological data offer high-resolution, direct visualization of fiber orientation in ex vivo tissue samples but are invasive and limited to non-living specimens. Conversely, diffusion tensor imaging allows for non-destructive, three-dimensional reconstruction of fiber fields in intact hearts, though it suffers from low spatial resolution, sensitivity to noise, and limited applicability in vivo due to long acquisition times and motion artifacts.

In this study, myocardial fiber orientation is prescribed using a rule-based method based on anatomical observations from histological data. The local myocardial coordinate system is defined pointwise to align with the anatomical structure of the ventricular wall. At each location, a right-handed orthonormal basis $[\hat{e}_t, \hat{e}_n, \hat{e}_l]$ is constructed. The unit vector $\hat{e}_t$ represents transmural direction obtained from the gradient of the wall distance field (i.e., the sheet-normal) as:

$$\hat{e}_t = \frac{\vec{\xi}}{\|\vec{\xi}\|} \tag{1}$$

Which $\vec{\xi}$ shows the transmural direction from the endocardium to the epicardium. Moreover, the $\hat{e}_n$ is normal direction, which can be defined based on apico-basal direction $k$ as follows [13]:

$$\hat{e}_n = \frac{k-(k.\hat{e}_t)\hat{e}_t}{\|k-(k.\hat{e}_t)\hat{e}_t\|} \tag{2}$$

Finally, the third unit vector $\hat{e}_l$, referred to as the longitudinal or circumferential direction, is computed as the cross-product between the previously defined orthonormal vectors $\hat{e}_n$ and $\hat{e}_t$:

$$\hat{e}_l = \hat{e}_t \times \hat{e}_n \tag{3}$$

This operation ensures a locally consistent, right-handed orthonormal frame at each point within the myocardial domain. As shown in Fig. 2-(a), the unit vector $\hat{e}_l$ and $\hat{e}_n$ lie tangential to the myocardial surface; however, the unit vector $\hat{e}_t$ is normal to the myocardial surface.

Following the construction of the local frame, the final fiber and sheet directions are computed using two sequential rotations derived from histological data [13]. The first rotation is applied about the transmural axis $\hat{e}_t$, which is shown by $\theta$. This operation adjusts the in-plane alignment of the fibers across the wall thickness and captures the experimentally observed helicoidal arrangement of myofibers. The maximum rotation angle around axis $\hat{e}_t$ at endocardium and epicardium of RV are represented by $\theta^{RV}_{endo,max}$, and $\theta^{RV}_{epi,max}$ as well as this angle at endocardium and epicardium of LV are denoted as $\theta^{LV}_{endo,max}$, and $\theta^{LV}_{epi,max}$, respectively. In literature [13], these rotation angles were assumed as

$$\theta^{RV}_{endo,max} = +90°,\ \theta^{RV}_{epi,max} = -25°, \quad \theta^{LV}_{endo,max} = +60°,\ \theta^{LV}_{epi,max} = -60° \tag{4}$$

The rotation angle $\theta$ is defined as a linear interpolation between endocardial and epicardial values according to the normalized transmural depth ($\beta$) as

$$\theta(\beta) = \theta_{epi}(1-\beta) + \theta_{endo}\,\beta \tag{5}$$

The dimensionless transmural depth $\beta$ is defined as

$$\beta = \frac{D_{epi}}{D_{epi} + D_{endo}} \tag{6}$$

where $D_{epi}$ and $D_{endo}$ show the shortest Euclidean distances to the epicardial and endocardial surfaces (of both RV and LV), respectively. The gradient of this scalar field defines the transmural direction, which is used as the first basis vector in a local coordinate system. The variables $\theta_{epi}$ and $\theta_{endo}$ blend left and right ventricular characteristics throughout the myocardium, as defined in Eq. (7).

$$\theta_{epi} = E^{RV}\,\theta^{RV}_{epi_{max}} + E^{LV}\,\theta^{LV}_{epi_{max}} \qquad \theta_{endo} = E^{RV}\,\theta^{RV}_{endo_{max}} + E^{LV}\,\theta^{LV}_{endo_{max}} \tag{7}$$

The weighted interpolation variables $E^{RV}$, and $E^{LV}$ are defined based on the nearest distance to RV's endocardium ($D^{RV}_{endo}$) and epicardium ($D^{RV}_{epi}$) as well as LV's endocardium ($D^{LV}_{endo}$) and epicardium ($D^{LV}_{epi}$) as follows:

$$E^{RV} = \frac{D^{LV}_{endo}}{D^{LV}_{endo} + D^{RV}_{endo}}, \quad E^{LV} = \frac{D^{RV}_{endo}}{D^{LV}_{endo} + D^{RV}_{endo}} \tag{8}$$

Through coordinate system transformation by rotation $\theta$ about the transmural axis $\hat{e}_t$, the rotated unit axis can be expressed as

$$[\hat{e}_t, \hat{e}_n, \hat{e}_l] \xrightarrow{\theta} [\hat{e}_t, \hat{e}'_n, \hat{e}'_l] \tag{9}$$

The histological data [13] also suggest a rotation $\alpha$ around the rotated axis $\hat{e}'_l$ with corresponding expressions similar to equations (5)-(8) and following angles:

$$\alpha^{RV}_{endo,max} = 0°,\ \alpha^{RV}_{epi,max} = -20°, \quad \alpha^{LV}_{endo,max} = +20°,\ \alpha^{LV}_{epi,max} = -20° \tag{10}$$

The local orthogonal fiber coordinate system is computed by rotation $\alpha$, as follows:

$$[\hat{e}_t, \hat{e}'_n, \hat{e}'_l] \xrightarrow{\alpha} [\mathbf{s}, \mathbf{n}, \mathbf{f}] \tag{11}$$

Which $\boldsymbol{s}$ is sheet unit vector, $\mathbf{n}$ is sheet-normal unit vector, and $\mathbf{f}$ shows the fiber direction. In Fig. 2(b), the local fiber coordinate system $[\mathbf{s}, \mathbf{n}, \mathbf{f}]$ is compared against the local myocardial coordinate system $[\hat{e}_t, \hat{e}_n, \hat{e}_l]$.

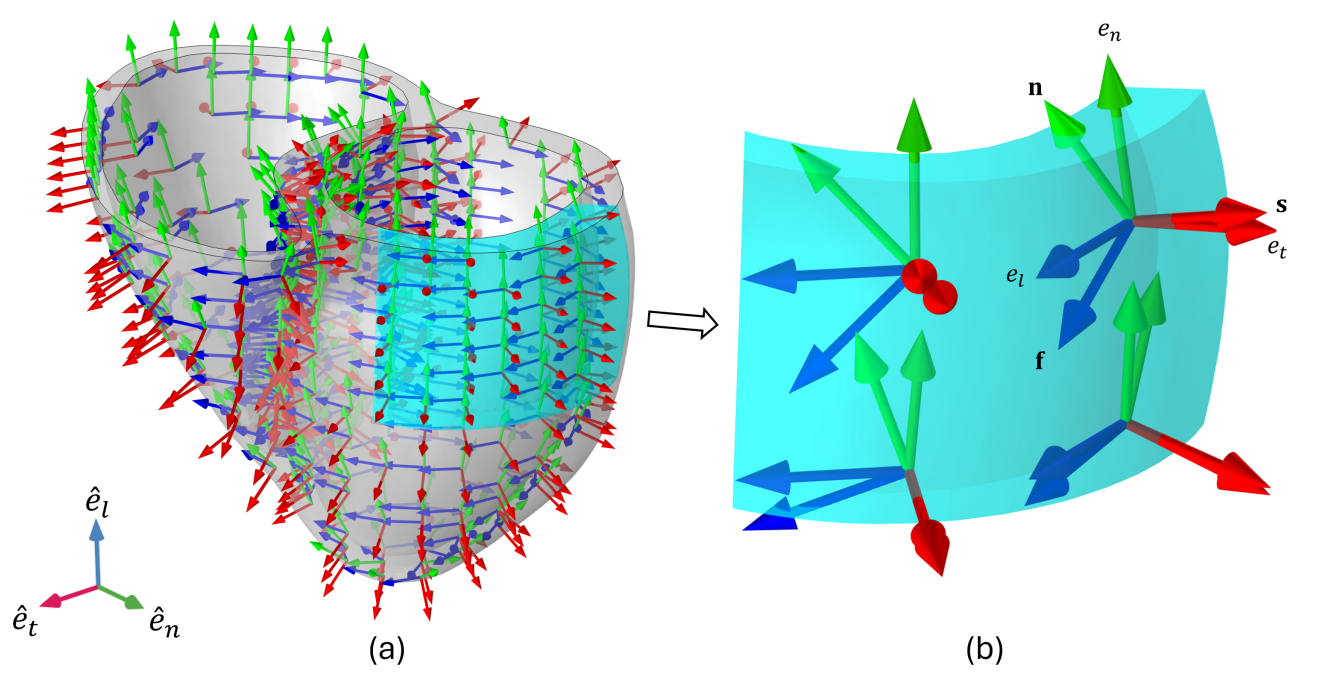

Fig. 2-(a) Local orthonormal coordinate system in the myocardium. The unit vectors $\hat{e}_t$ , $\hat{e}_n$, and $\hat{e}_l$ represent the transmural (sheet-normal), apico-basal (normal), and longitudinal (circumferential) directions, respectively; (b) the comparison of the local fiber coordinate system $[\mathbf{s}, \mathbf{n}, \mathbf{f}]$ against local myocardial coordinate system $[\hat{e}_t, \hat{e}_n, \hat{e}_l]$.

Fig. 3 represents the helical distribution of myofiber over the epicardium and endocardium along with the local fiber coordinate system. The rotation angle $\theta$ is also shown by color map, indicating the orientation -60° and +60° at LV's epicardium and endocardium as well as the orientation +90° and -25° at RV's epicardium and endocardium, respectively, which is consistent with Eq. (4). Moreover, the zoom-box confirms the fiber direction matches with direction of the unit vector $\mathbf{f}$.

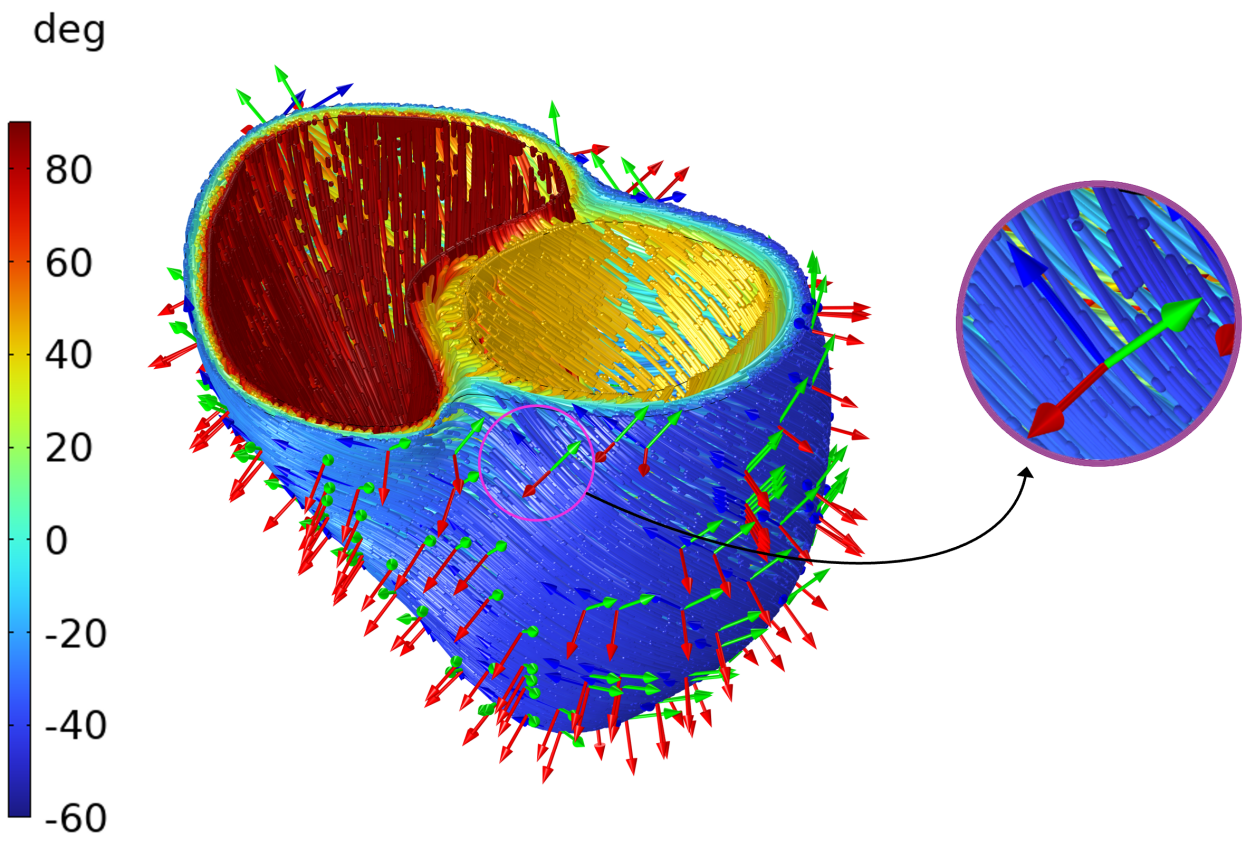

Fig. 3- Helical distribution of myocardial fibers across endocardial and epicardial surfaces of the biventricular model. The color map represents the transmural fiber angle θ, varying from +60° to –60° in the LV and from +90° to –25° in the RV. Overlaid vectors denote the local orthonormal fiber coordinate system, with a zoomed-in view confirming alignment of the computed fiber direction $\mathbf{f}$ with anatomical orientation.

Fig. 4 represents four different visualizations of the biventricular heart model, illustrating the distribution of key variables used in the electromechanical model. The top two images, (a) and (b), show the weighted interpolation variables, $E^{RV}$ and $E^{LV}$ respectively. As previously defined, the value of $E^{RV}$ and $E^{LV}$ approaches to 1 at regions close to the right and left ventricles, respectively.

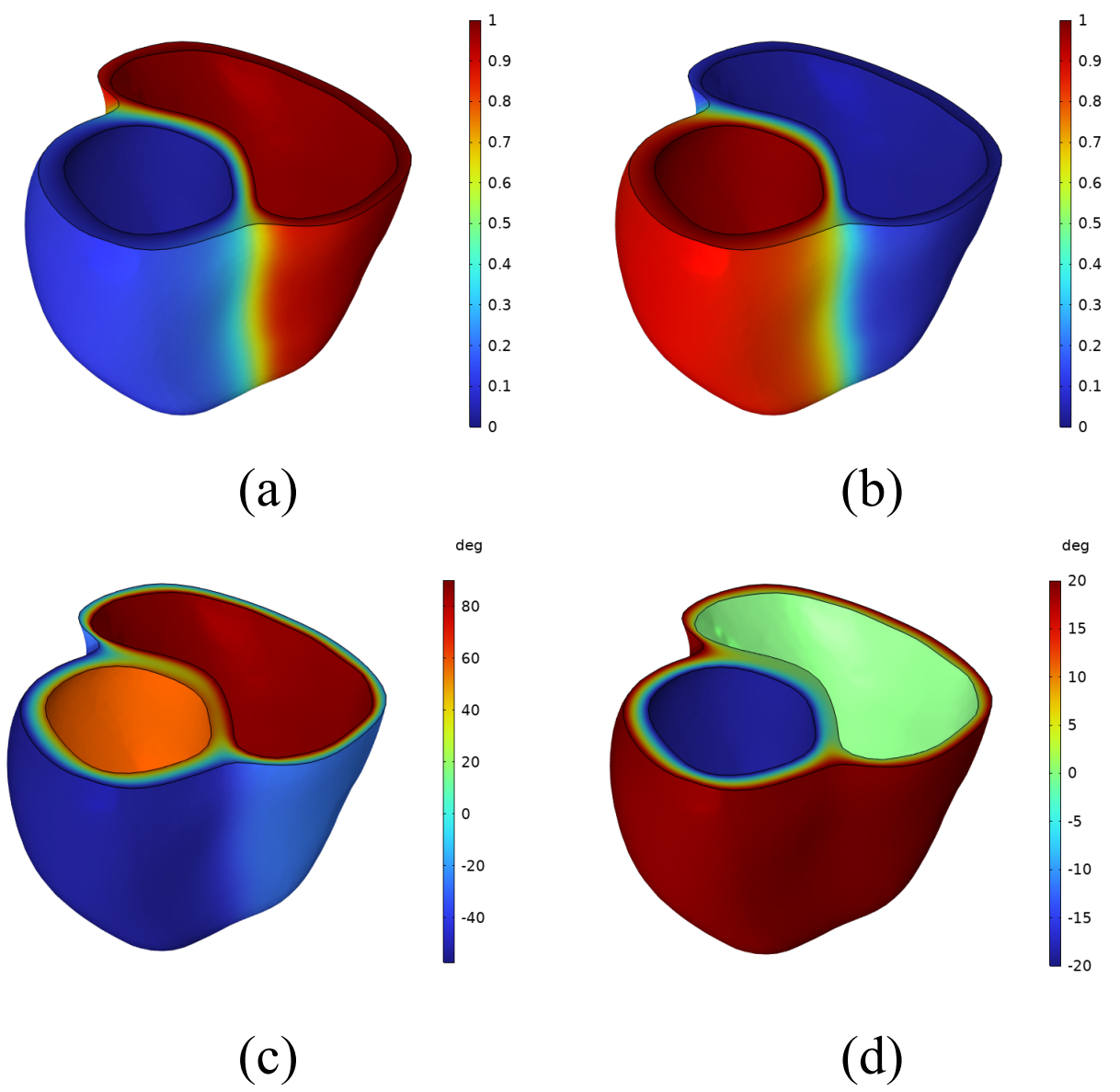

Fig. 4- The distribution of the weighted interpolation variables (a) $E^{RV}$, and (b) $E^{LV}$ as well as the interpolated rotation angle (c) $\theta$, and (d) $\alpha$.

## C. Electrophysiology

Electrophysiological activity within the myocardium is modeled using a reaction-diffusion framework that describes the spatiotemporal propagation of the transmembrane potential. The monodomain equation governs this process and is expressed as [19]:

$$\chi_m C_m \frac{\partial \varphi}{\partial t} - \nabla.(D \nabla \varphi) + \chi_m I_{ion}(\varphi, F, r) = 0 \tag{12}$$

where $\varphi$ denotes the transmembrane electric potential, $\chi_m$ the membrane surface-to-volume ratio, $C_m$ the membrane capacitance, $D$ the anisotropic conductivity tensor, and $I_{ion}$ the total ionic current density per unit membrane area. To capture physiologically realistic conduction velocities, the fiber-aligned anisotropic conductivity tensor $D$ is defined based on the coefficients of isotropic conductivity $d_{iso}$ and anisotropic conductivity $d_{ani}$ along fibers, as given by:

$$D = d_{iso} C_m \chi_m \mathbf{I} + d_{ani} C_m \chi_m \mathbf{f} \otimes \mathbf{f} \tag{13}$$

The ionic current comprises two principal dimensionless components: an excitation-induced term $\tilde{I}_e(\varphi, r)$ and a stretch-induced term $\tilde{I}_m(F)$, such that:

$$I_{ion}(\varphi, F, r) = C_m \frac{\beta_\phi}{\beta_t} \left( \tilde{I}_e(\varphi, r) + \tilde{I}_m(F) \right) \tag{14}$$

The variable $F$ represents the deformation gradient tensor.

Here, $C_m$ is the membrane capacitance, $\beta_\phi$ is the voltage-scaling factor mapping the dimensionless potential $\phi$ to the dimensional potential $\varphi$, $\beta_t$ is the (activation-time-dependent) time-scaling parameter mapping the dimensionless time $\tau$ to physical time t, and F is the deformation gradient, with the following equations.

$$t = \beta_t \tau \tag{15}$$

$$\phi = (\varphi - \delta_\phi)/\beta_\phi \tag{16}$$

Indeed, the transmembrane potential is rescaled such that the resting and peak values correspond to -80 mV and 20 mV, respectively. Temporal scaling was adjusted according to spatial location, particularly the distance from the apex along the Z-direction, to replicate apex-to-base repolarization gradients observed in vivo.

Moreover, $r$ is a recovery variable, as:

$$\frac{\partial r}{\partial \tau} = \left(\gamma + \frac{\mu_1}{\mu_2 + \phi} r\right)\left(-r - c\phi(\phi - b - 1)\right) \tag{17}$$

The Aliev–Panfilov (AP) model is adopted to model the excitation-induced term $I_e$ [31]. This phenomenological model captures the nonlinear dynamics of depolarization and repolarization through a system of dimensionless ordinary differential equations:

$$\tilde{I}_e(\varphi, r) = c\phi(\phi - a)(\phi - 1) + r\phi \tag{18}$$

where $\phi$ is the dimensionless electric potential, as follows:

$$\phi = (\varphi - \delta_\phi)/\beta_\phi \tag{19}$$

In addition, the mechanical deformation feedback is introduced via the stretch-induced current term $I_m$, defined by:

$$\tilde{I}_m = \Theta G_s(\lambda - 1)(\phi - \phi_s) \tag{20}$$

where $\lambda$ is the fiber stretch, $G_s$ the maximum conductance, $\phi_s$ the resting potential of the stretched state, and $\Theta$ is an activation parameter that modulates the stretch contribution. This formulation enables electromechanical coupling in the model by relating electrical activation to tissue deformation. This variable $\Theta$ equals 1 when fibers are stretched and 0 when they are compressed.

The atrioventricular (AV) plane, composed of a dense ring of fibrous tissue, acts as a crucial electrical insulator separating the atria from the ventricles. This region is non-conductive to electrical propagation, which is shown in Fig. 5(a) with green color. Consistent with this anatomy, the monodomain model solves electrophysiology equations exclusively within the conductive ventricular myocardium (red region in Fig. 5(a)). This is enforced by applying an insulation (zero-flux) boundary condition across the ventricular base. Similarly, the zero-flux (Neumann) boundary condition is assigned on all external surfaces. An initial depolarizing stimulus of -10 mV was applied at the base of the interventricular septum, mimicking excitation from the AV node, as shown in Fig. 5. The resulting excitation propagates throughout the ventricular myocardium and triggers spatially heterogeneous contractions. The model accounts for the apex-to-base gradient in repolarization by employing a time-scaling function that modulates activation duration based on spatial coordinates, thereby reproducing observed mechanical restitution patterns.

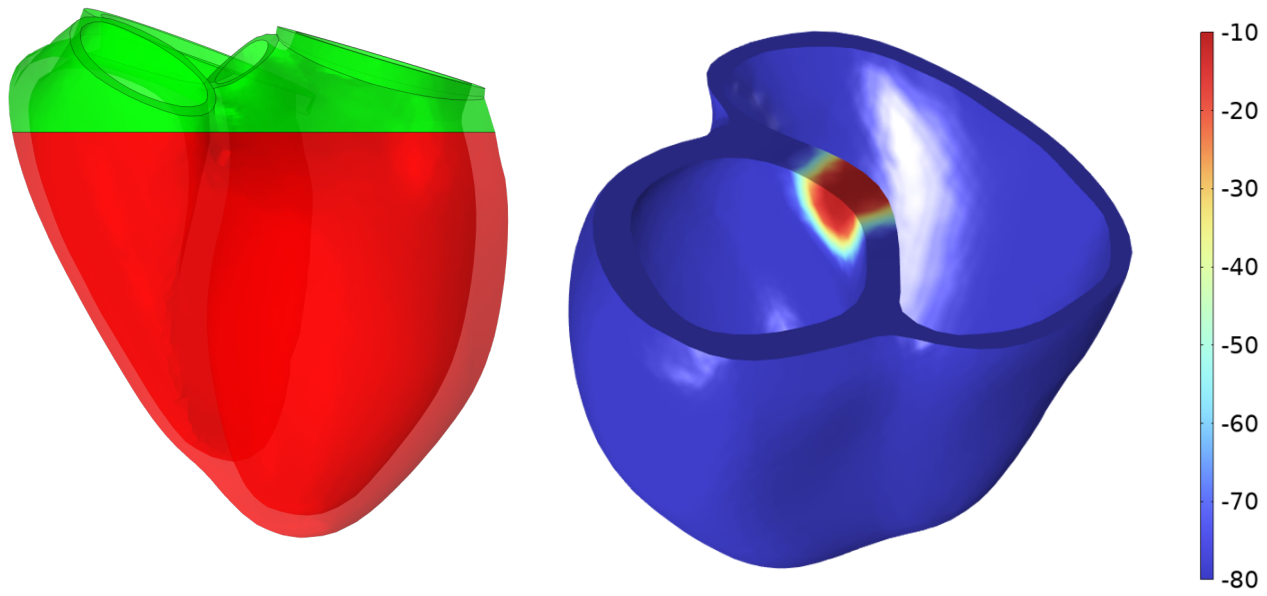

Fig. 5- (a) The non-conductive myocardium region close to the AV plane (green) separated from the conductive myocardium region (red), (b) the initial depolarizing stimulus of -10 mV applied at the base of the interventricular septum.

The considered values of the electrophysiology parameters used in this study are presented in Appendix A (Table A1).

## D. Electromechanical Activation

The link between electrical excitation and mechanical contraction is modeled by incorporating a voltage-dependent active stress term into the hyperelastic framework. Upon depolarization, the local transmembrane potential triggers an internal stress response that evolves dynamically over time. The active stress $s_a$, which arises in response to electrical activation, is incorporated additively into the second Piola–Kirchhoff stress tensor and contributes to the total mechanical response of the tissue.

The evolution of $s_a$ is governed by a first-order differential equation:

$$\frac{\partial s_a}{\partial t} = \varepsilon(\varphi)(k(\varphi - \varphi_r) - s_a) \tag{21}$$

where $\varphi_r$ is the resting transmembrane potential (-80 mV), and $\varepsilon(\varphi)$ represents a voltage-dependent delay function defined as [31]:

$$\varepsilon(\varphi) = \varepsilon_0 + (\varepsilon_0 - \varepsilon_1)e^{-e^{-\vartheta(\varphi - \varphi_t)}} \tag{22}$$

Here, $k$ denotes the stress generation rate, and $\varepsilon_0$, $\varepsilon_1$ , $\vartheta$ , $\varphi_t$ are empirically calibrated parameters that control the timing and amplitude of the active stress response. The generated active tension is projected onto the orthotropic myocardial structure, contributing to the stress tensor along the fiber (**f**), sheet (**s**), and sheet-normal (**n**) directions:

$$S_{act} = s_a(v_f \mathbf{f} \otimes \mathbf{f} + v_s \mathbf{s} \otimes \mathbf{s} + v_n \mathbf{n} \otimes \mathbf{n}) \tag{23}$$

The dimensionless coefficients $v_f$, $v_s$, and $v_n$ are weighting factors that define the proportion of active stress aligned with each anatomical axis. The values of active stress parameters are expressed in Appendix A (Table A1).

This formulation ensures that active contraction is anisotropically distributed, consistent with the local fiber architecture. The overall mechanical behavior of the tissue arises from the combination of this active component and the passive hyperelastic response described previously.

## *E. 0D closed-loop Hemodynamic and Atria Models*

The 0D closed-loop hemodynamic model provides a lumped-parameter representation of the cardiovascular system, integrating systemic and pulmonary circulations with cardiac chamber dynamics and valve behavior [18], [19]. In this framework, vascular segments are modeled via electrical analogs comprising resistances ($\mathbb{R}$), inductances ($\mathbb{L}$), and capacitances ($\mathbb{C}$), representing viscous resistance of blood, inertia of blood flow, and elasticity of blood vessels, respectively. The coupling between the 3D biventricular model and the 0D model is shown in Fig. 6.

Throughout this paper, hemodynamic variables in 0D model are named $X^{SYS/PUL}_{AR/VEN}$, where $X$ is the physical quantity ($p$: blood pressure, $Q$: blood flow rate), the superscripts SYS and PUL specify whether it belongs to the systemic or pulmonary circulation, and the subscripts AR and VEN determine whether it refers to the arterial or venous compartment, respectively. Moreover, the related variables in 3D model are $Y_{RV/LV}$, where $Y$ is the physical quantity ($V$: volume, $p$: pressure), the subscripts RV and LV show that this parameter belongs to the right or left ventricle, respectively.

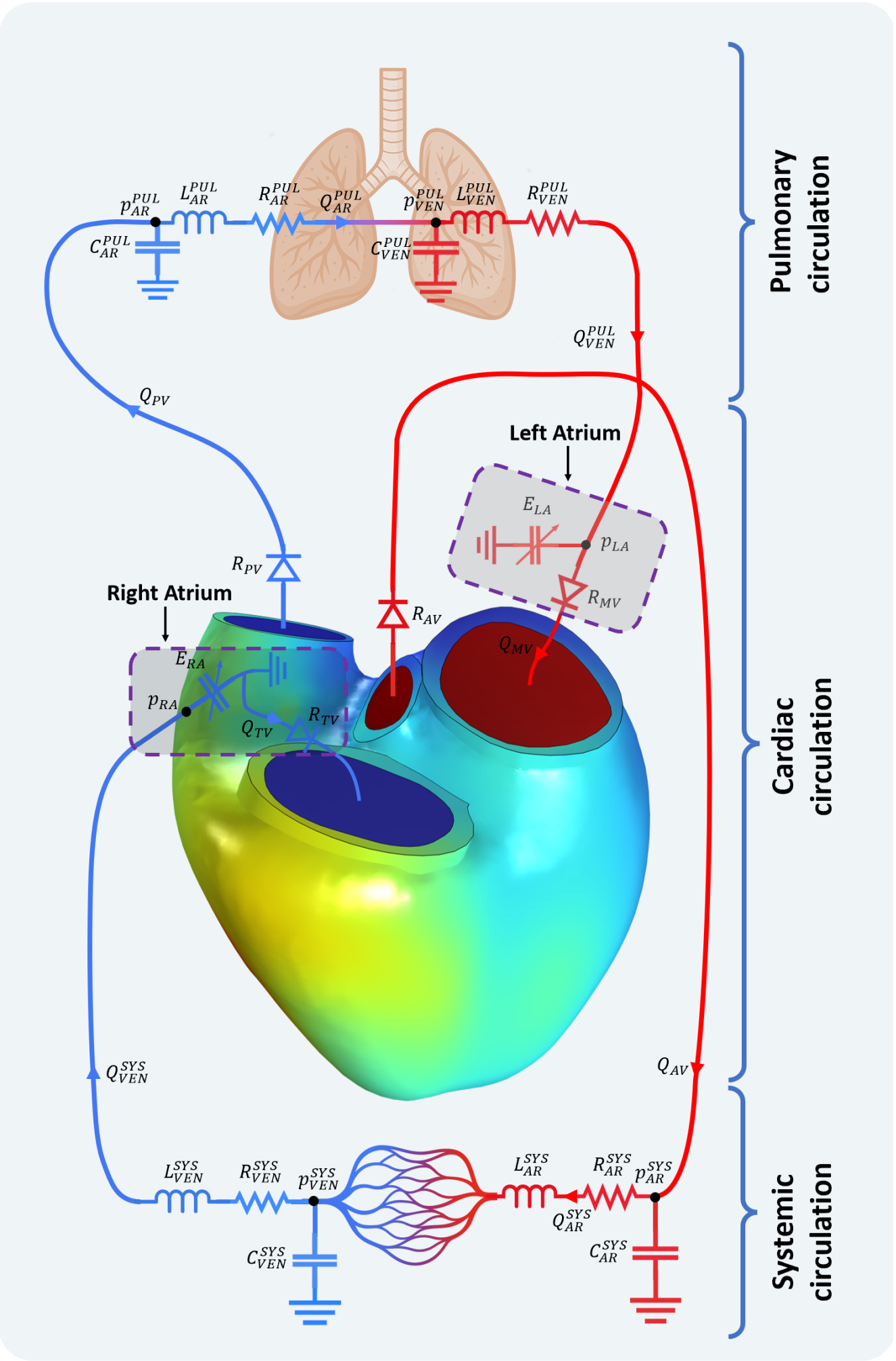

Fig. 6- The 3D-0D coupling between the 3D biventricular model and the 0D hemodynamic and atria model.

In the following, the governing equations for cardiac chamber volume, arterial-venous pressure, and arterial-venous flow rate dynamics are presented using fundamental principles of fluid mechanics and mass conservation.

- Cardiac chamber volume dynamics:

The cardiac chamber volume dynamics can be expressed based on mass conservation for each chamber. Therefore, the time-rate of change of its volume equals inflow minus outflow, as Eq. (24).

$$\begin{aligned}\dot{V}_{LA}(t) &= Q^{PUL}_{VEN}(t) - Q_{MV}(t)\\ \dot{V}_{RA}(t) &= Q^{SYS}_{VEN}(t) - Q_{TV}(t)\\ \dot{V}_{LV}(t) &= Q_{MV}(t) - Q_{AV}(t)\\ \dot{V}_{RV}(t) &= Q_{TV}(t) - Q_{PV}(t)\end{aligned} \tag{24}$$

- Arterial–venous pressure dynamics:

The time derivative of the arterial or venous reservoir's pressure equals the net inflow minus outflow scaled by the compartment's compliance.

$$\begin{aligned}\dot{p}^{SYS}_{AR}(t) &= \left(Q_{AV}(t) - Q^{SYS}_{AR}(t)\right)/\mathbb{C}^{SYS}_{AR}\\ \dot{p}^{PUL}_{AR}(t) &= \left(Q_{PV}(t) - Q^{PUL}_{AR}(t)\right)/\mathbb{C}^{PUL}_{AR}\\ \dot{p}^{SYS}_{VEN}(t) &= \left(Q^{SYS}_{AR}(t) - Q^{SYS}_{VEN}(t)\right)/\mathbb{C}^{SYS}_{VEN}\\ \dot{p}^{PUL}_{VEN}(t) &= \left(Q^{PUL}_{AR}(t) - Q^{PUL}_{VEN}(t)\right)/\mathbb{C}^{PUL}_{VEN}\end{aligned} \tag{25}$$

- Arterial–venous flow rate:

The time derivative of arterial or venous flow rates can be defined as:

$$\begin{aligned}\dot{Q}^{SYS}_{AR}(t) &= \left(-Q^{SYS}_{AR}(t) - \left(p^{SYS}_{VEN}(t) - p^{SYS}_{AR}(t)\right)/\mathbb{R}^{SYS}_{AR}\right)\mathbb{R}^{SYS}_{AR}/\mathbb{L}^{SYS}_{AR}\\ \dot{Q}^{SYS}_{VEN}(t) &= \left(-Q^{SYS}_{VEN}(t) - \left(p_{RA}(t) - p^{SYS}_{VEN}(t)\right)/\mathbb{R}^{SYS}_{VEN}\right)\mathbb{R}^{SYS}_{VEN}/\mathbb{L}^{SYS}_{VEN}\\ \dot{Q}^{PUL}_{AR}(t) &= \left(-Q^{PUL}_{AR}(t) - \left(p^{PUL}_{VEN}(t) - p^{PUL}_{AR}(t)\right)/\mathbb{R}^{PUL}_{AR}\right)\mathbb{R}^{PUL}_{AR}/\mathbb{L}^{PUL}_{AR}\\ \dot{Q}^{PUL}_{VEN}(t) &= \left(-Q^{PUL}_{VEN}(t) - \left(p_{LA}(t) - p^{PUL}_{VEN}(t)\right)/\mathbb{R}^{PUL}_{VEN}\right)\mathbb{R}^{PUL}_{VEN}/\mathbb{L}^{PUL}_{VEN}\end{aligned} \tag{26}$$

Moreover, the four cardiac valves are modeled as a non-ideal diode. The parameters $\mathbb{R}_{PV}$, $\mathbb{R}_{AV}$, $\mathbb{R}_{MV}$, and $\mathbb{R}_{TV}$ denote the valve resistances, switching between a low value $R_O$ during the open phase to account for the forward-flow pressure drop and a high value $R_C$ that represents the near-blocking resistance during valve closure. The valve resistance $\mathbb{R}_i$ ($i \in \{MV, AV, TV, PV\}$) is defined piece-wise as in Eq. (27) to ensure a smooth transition between the close and open phases.

$$\mathbb{R}_i(\Delta p) = \begin{cases} R_O & \Delta p > \Delta p_{tran} & \text{(open phase)}\\ R_{tran} & -\Delta p_{tran} \le \Delta p \le \Delta p_{tran} & \text{(transient phase)}\\ R_C & \Delta p < -\Delta p_{tran} & \text{(close phase)}\end{cases} \tag{27}$$

Here, Δp is the trans-valvular pressure difference defined with the forward (physiologic) flow direction. Therefore, the blood flow at the mitral, aortic, tricuspid, and pulmonary valves is presented as $Q_{MV}$, $Q_{AV}$, $Q_{TV}$, and $Q_{PV}$, respectively. These volumetric blood flows can be expressed as follows.

$$
\begin{aligned}
Q_{MV}(t) &= (p_{LA}(t) - p_{LV}(t))/\mathbb{R}_{MV}(p_{LA}(t) - p_{LV}(t)) \\
Q_{AV}(t) &= (p_{LV}(t) - p_{AR}^{SYS}(t))/\mathbb{R}_{AV}(p_{LV}(t) - p_{AR}^{SYS}(t)) \\
Q_{TV}(t) &= (p_{RA}(t) - p_{RV}(t))/\mathbb{R}_{TV}(p_{RA}(t) - p_{RV}(t)) \\
Q_{PV}(t) &= (p_{RV}(t) - p_{AR}^{PUL}(t))/\mathbb{R}_{PV}(p_{RV}(t) - p_{AR}^{PUL}(t))
\end{aligned} \tag{28}
$$

As shown in Fig. 6, the atrial chambers are represented by time-varying elastance functions of $EL_{LA}(t)$ for the left atrium and $EL_{RA}(t)$ for the right. The expression in Eq. (29) models the pressure (p) of the left and right atria as a function of their elastance $(EL)$ and their volume $(V)$ relative to a reference volume $(V_0)$, which are detailed in [19].

$$
\begin{aligned}
p_{LA}(t) &= EL_{LA}(t)(V_{LA} - V_{LA,0}) \\
p_{RA}(t) &= EL_{RA}(t)(V_{RA} - V_{RA,0})
\end{aligned} \tag{29}
$$

The values of defined hemodynamic constants in the 0D model are expressed in Appendix A (Table A2).

## F. Two-way Fluid-Structure-Interaction

Blood flow in the ventricular chambers is coupled bidirectionally to the myocardium (at the endocardium surface) via a two-way FSI formulation implemented in COMSOL Multiphysics. The solid mechanics of the biventricular wall with the electromechanical activation (described in section 2.4) is solved in a Lagrangian frame, while the blood is modeled as an incompressible fluid in an Arbitrary-Lagrangian–Eulerian (ALE) frame on a deforming mesh that conforms to the endocardium. The coupling transmits wall motion to the fluid and, conversely, the unsteady fluid pressure and viscous tractions back to the myocardium. The 3D fluidic fields are also linked to the 0D closed-loop hemodynamics (described in section 2.5) at the four valve annuli (Fig. 6), completing a 3D-0D-electromechanical closed loop.

In literature, blood cardiac chambers are modeled as a viscous, incompressible, and Newtonian fluid [1]. Therefore, the constant density $\rho_{fluid}$ and viscosity $\mu_{fluid}$ are attributed to blood flow in cardiac cavities. In the dynamic fluid domain $\Omega_f(t)$, the incompressible Navier–Stokes equations are implemented with Cauchy stress $\sigma_{fluid} = -p\mathbf{I} + 2\,\mu_{fluid}\,\varepsilon(v_{fluid})$ and mesh velocity $v_{mesh} = \partial u_{mesh}/\partial t$ obtained from the moving-mesh displacement $u_{mesh}$. The strain-rate tensor $\varepsilon(v_{fluid})$ equals $(\nabla v_{fluid} + \nabla v_{fluid}^T)/2$. In this study, the intraventricular blood flow is modeled as laminar (i.e., no explicit turbulence model) to improve numerical robustness and reduce computational cost in the two-way ALE FSI; as our focus is on endocardial wall motion, transitional effects near valve jets are neglected.

In the solid (myocardial) domain $\Omega_s$ the momentum balance is written in the reference configuration, as follows:

$$
\rho_{solid}\frac{\partial^2 u_{solid}}{\partial t^2} = \nabla.\sigma_{solid} \qquad \text{, in } \Omega_f(t) \tag{30}
$$

with Cauchy stress $\sigma_{solid}$ obtained from the passive hyperelastic law plus the active stress of Eq. (23).

At the endocardial fluid–structure interface $\Gamma_{fs}(t)$, two conditions ensure tight two-way coupling. Kinematic continuity enforces no slip and no penetration by matching the fluid velocity to the wall velocity: $v_{fluid} = \dot{u}_{solid}$ on $\Gamma_{fs}(t)$, which also sets the ALE mesh velocity $v_{mesh} = v_{fluid} = \dot{u}_{solid}$, on the interface. Dynamic equilibrium balances tractions across the interface: $\sigma_{fluid}n = \sigma_{solid}n$. In COMSOL, the FSI, ALE Multiphysics coupling enforces these relations each time step, mapping endocardial displacement/velocity to the fluid mesh while applying computed fluid tractions back to the solid.

## G. Mechanical Boundary Conditions for Neighboring Tissues

The ventricles are enclosed by the serous and fibrous pericardium and interface with adjacent thoracic structures: the diaphragm, mediastinal fascia, and lungs via the pleura. The fibrous pericardium is a tough outer sac, while the serous pericardium is a double-layered membrane with a lubricating fluid in between. At the base, the myocardium is tethered to the fibrous skeleton and atria and great vessels. These neighboring tissues can significantly affect the heart's motion. In modeling, these influences are efficiently represented as surface constraints on the epicardium and basal cut planes.

All boundary tractions are expressed in the local myocardial coordinate system with the orthonormal basis $[\hat{e}_t, \hat{e}_n, \hat{e}_l]$. The epicardial surface $(\Gamma_{epi})$ and the basal plane $(\Gamma_{basal})$ are located in conductive and non-conductive myocardium regions (shown in Fig. 5(a)), respectively, which have distinct physiological roles that necessitate different boundary conditions within the model. The viscoelastic tractions on epicardial and basal surfaces $(\boldsymbol{t}^{\Gamma})$ are defined as

$$
\boldsymbol{t}^{\Gamma} = (k_t^{\Gamma}P_t + k_n^{\Gamma}P_n + k_l^{\Gamma}P_l)\boldsymbol{u}_{solid} + (c_t^{\Gamma}P_t + c_n^{\Gamma}P_n + c_l^{\Gamma}P_l)\dot{\boldsymbol{u}}_{solid} \quad \text{, on } \Gamma = \Gamma_{epi}, \Gamma_{basal} \tag{31}
$$

The $P_{\mathbb{i}} = \hat{e}_{\mathbb{i}} \times \hat{e}_{\mathbb{i}}$, $k_{\mathbb{i}}^{\Gamma}$ and $c_{\mathbb{i}}^{\Gamma}$ for $\mathbb{i} \in \{t, n, l\}$ are the corresponding directional projectors, stiffness, and damping coefficients, respectively. Usually, the apex exhibits minimal excursion over the cardiac cycle [32]. To suppress non-physiological drift, the epicardial restraint near the apex is reinforced by ramping the directional stiffness and damping with the normalized apico-basal coordinate $Z$ (with $Z = 0$ at the base; $Z_{cr}$= -5cm marks the cutoff defining the apical reinforcement zone for $Z \le Z_{cr}$), as given by

$$
k_{\mathbb{i}}^{\Gamma_{epi}}(Z) = \begin{cases} K_{\mathbb{i}}^{\Gamma_{epi}} & Z > Z_{cr} \\ K_{\mathbb{i}}^{\Gamma_{epi}}(1 + 10(Z_{cr} - Z)) & Z \le Z_{cr} \end{cases}, \mathbb{i} \in \{t, n, l\} \tag{32}
$$

$$
c_{\mathbb{i}}^{\Gamma_{epi}}(Z) = \begin{cases} C_{\mathbb{i}}^{\Gamma_{epi}} & Z > Z_{cr} \\ C_{\mathbb{i}}^{\Gamma_{epi}}(1 + 2(Z_{cr} - Z)) & Z \le Z_{cr} \end{cases}, \mathbb{i} \in \{t, n, l\} \tag{33}
$$

The spring/elastic boundary condition is not applied at the basal region $\Gamma_{basal}$ to preserve physiological AV-plane excursion and long-axis shortening. Instead, a damping-only boundary is used to mimic the compliant, energy-dissipating tether provided by the atria and the AV fibrous skeleton: it supplies viscous resistance (and thus suppresses non-physiological high-frequency vibrations) without introducing a restorative elastic force that would artificially restrain the base or reduce AV-plane displacement. In this way, the damper acts as a surrogate for atrial coupling, stabilizing the basal motion while leaving the natural base–apex shortening unbiased.

The values of boundary conditions coefficients are in Appendix C (Table A2).

## III. Result

### I. Cardiac Function Simulation

This section presents the outputs of the electromechanical heart model at a heart rate of 75 bpm. A cross-sectional representation of the model's state at key moments during the systolic and diastolic phases is shown in Fig. 7. The figure uses color mapping to indicate electrophysiological activation within the myocardium and blood flow velocity within the chambers. Furthermore, vectors are used to illustrate the unit vector of blood flow direction on the cross-sectional surfaces.

Systole begins in Fig. 7(a), taken at t=0 ms. The image shows the initial electrophysiological signal starting near the atrioventricular node. This corresponds to the isovolumetric contraction phase, where the heart's volume remains constant because all four valves are closed. At t=100 ms, Fig. 7(b) shows the heart during systole. The electrical activation wave has spread throughout the myocardium, and the myocardium begins to contract. This contraction significantly increases ventricular pressure, eventually exceeding the pressure in the great arteries and forcing the semilunar valves to open. As a result, blood is forcefully ejected from the ventricles.

Fig. 7(c), at t=250 ms, marks the end of systole. The myocardium is now relaxing, and ventricular pressure is beginning to decrease. When this pressure drops below the arterial pressure, the semilunar valves close to prevent blood from flowing back into the ventricles. The diastole phase is depicted in Fig. 7(d), taken at t=500 ms. As the ventricles continue to relax, their pressure falls below the pressure in the atria, causing the atrioventricular valves to open. This allows blood to passively flow from the atria into the ventricles, preparing the heart for the next cardiac cycle.

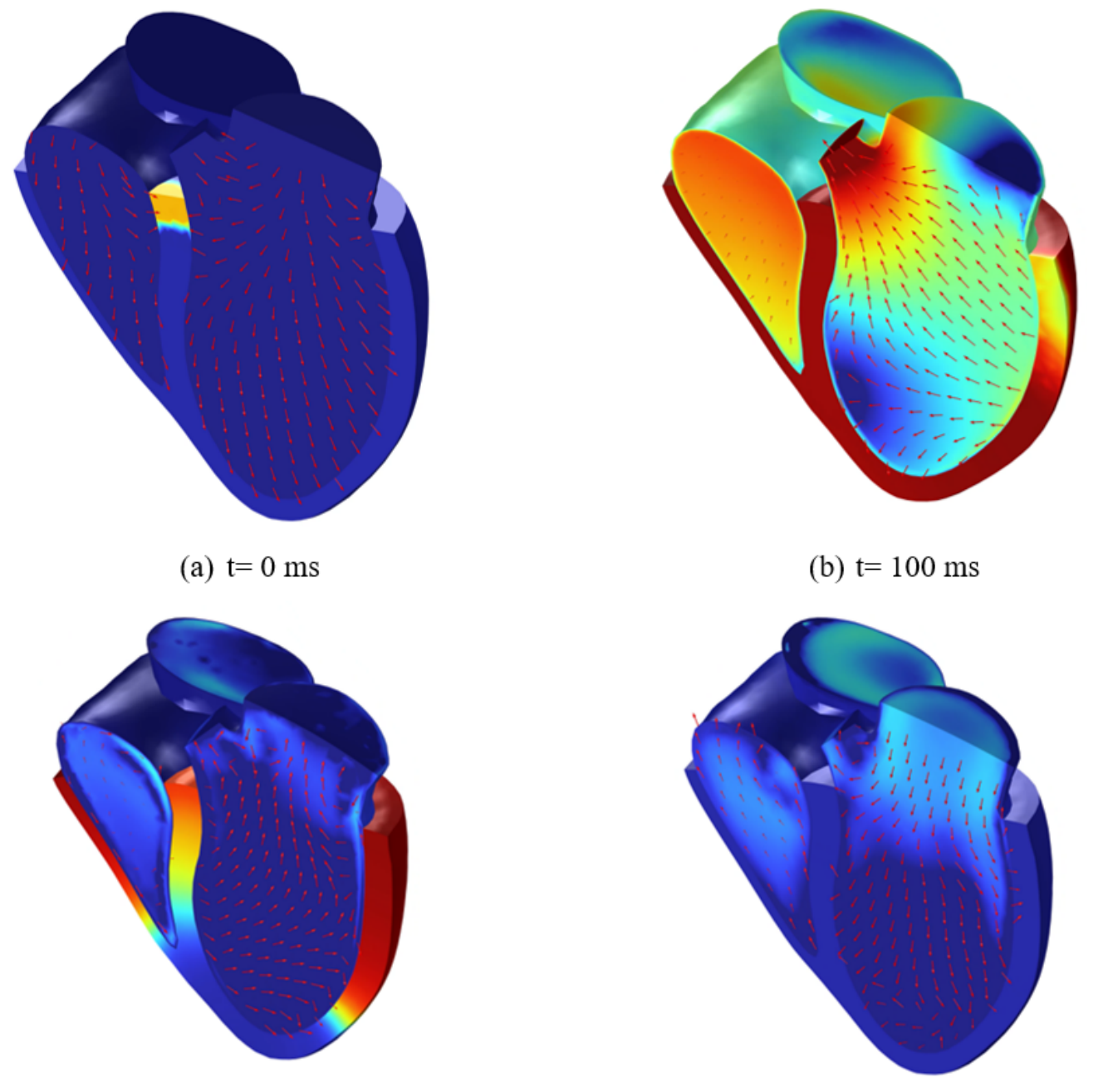

(a) t= 0 ms (b) t= 100 ms (c) t= 250 ms (d) t= 500 ms

Fig. 7- Cross-sectional view of the electromechanical model at various time points during the cardiac cycle. The color mapping in the myocardium indicates the electrophysiology level, while the color mapping in the chamber volumes represents blood flow velocity. The vectors illustrate the unit vector of blood flow velocity. Panel (a) shows the start of systole, (b) shows the heart during peak systole with near-complete myocardial activation, (c) shows the end of systole, and (d) shows a moment during diastole.

### II. Cine-MRI Heart Motion Evaluation

To validate the regional motion predictions of the developed digital twin framework, the electromechanical simulation results are compared with Cine MRI-derived motion from a healthy human dataset obtained from the AMRG Cardiac Atlas [33]. This dataset contains 13 SAX and 12 LAX slices acquired over one cardiac cycle, with 32 frames per cycle and a heart rate of approximately 67 bpm

Accurate validation of regional cardiac motion requires an imaging plane that can clearly capture ventricular deformation throughout the cardiac cycle. In this study, the long-axis Cine-MRI view is selected because it provides continuous visualization of ventricular motion from the basal to the apical regions, allowing evaluation of the RV's dominant longitudinal displacement. Among the 12 LAX views available in the dataset, all include the LV chamber, while only 7 of these LAX views capture the RV chamber, as shown in Fig. 8. In these LAX images, the LV endocardial segmentation is shown in red, whereas the RV endocardial segmentation is shown in cyan/blue. Therefore, only the slices containing the cyan/blue RV contour were selected for subsequent RV motion evaluation.

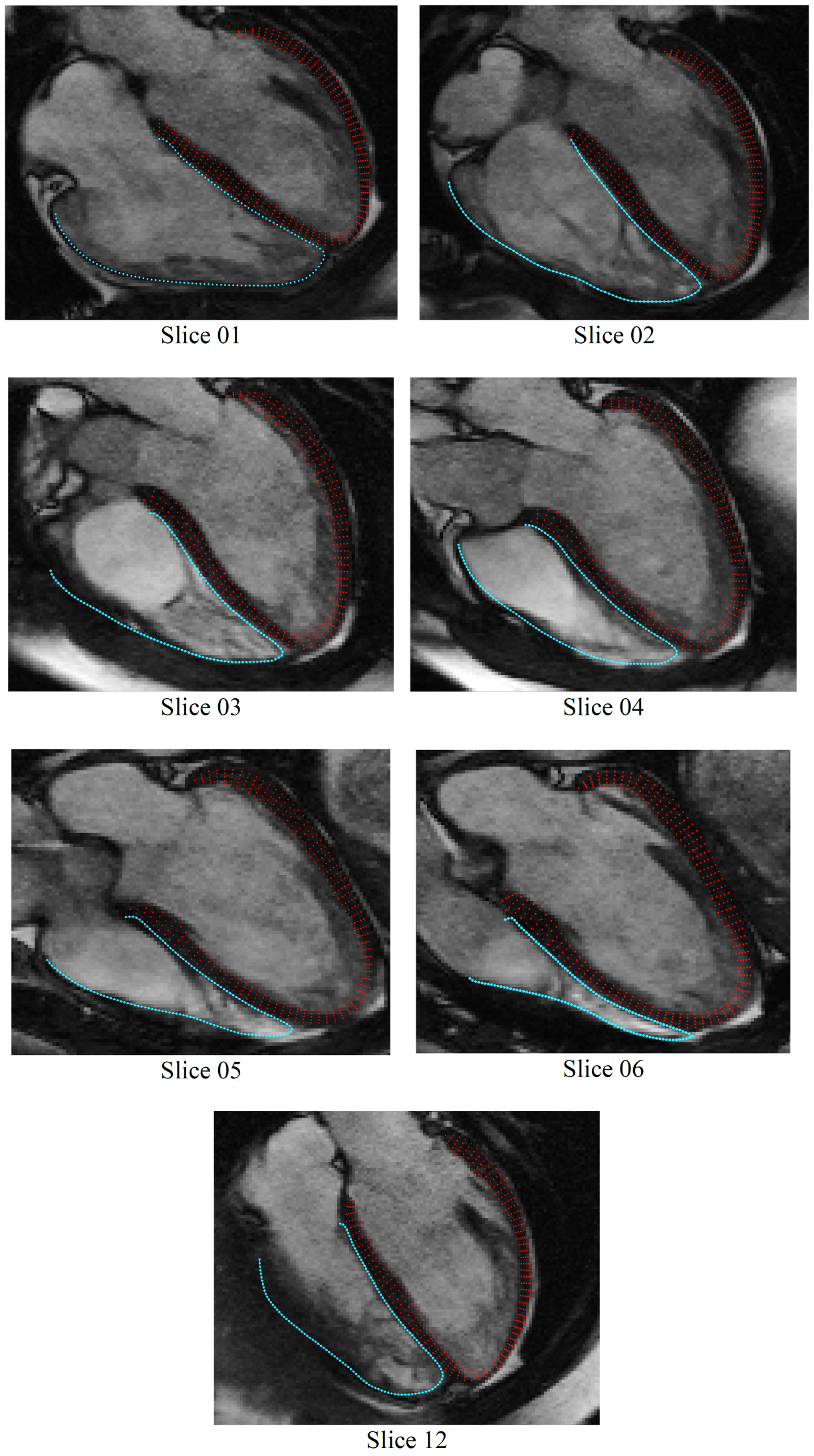

Fig. 8- Long-axis Cine-MRI slices containing the RV chamber used for RV motion evaluation.

Sequential frames of these slices are processed through segmentation and feature tracking by Segment software [34] to derive the displacement and motion pattern of the RV endocardium over one cardiac cycle. For instance, Fig. 9 illustrates the segmentation and feature tracking of slice 01 during one cardiac cycle.

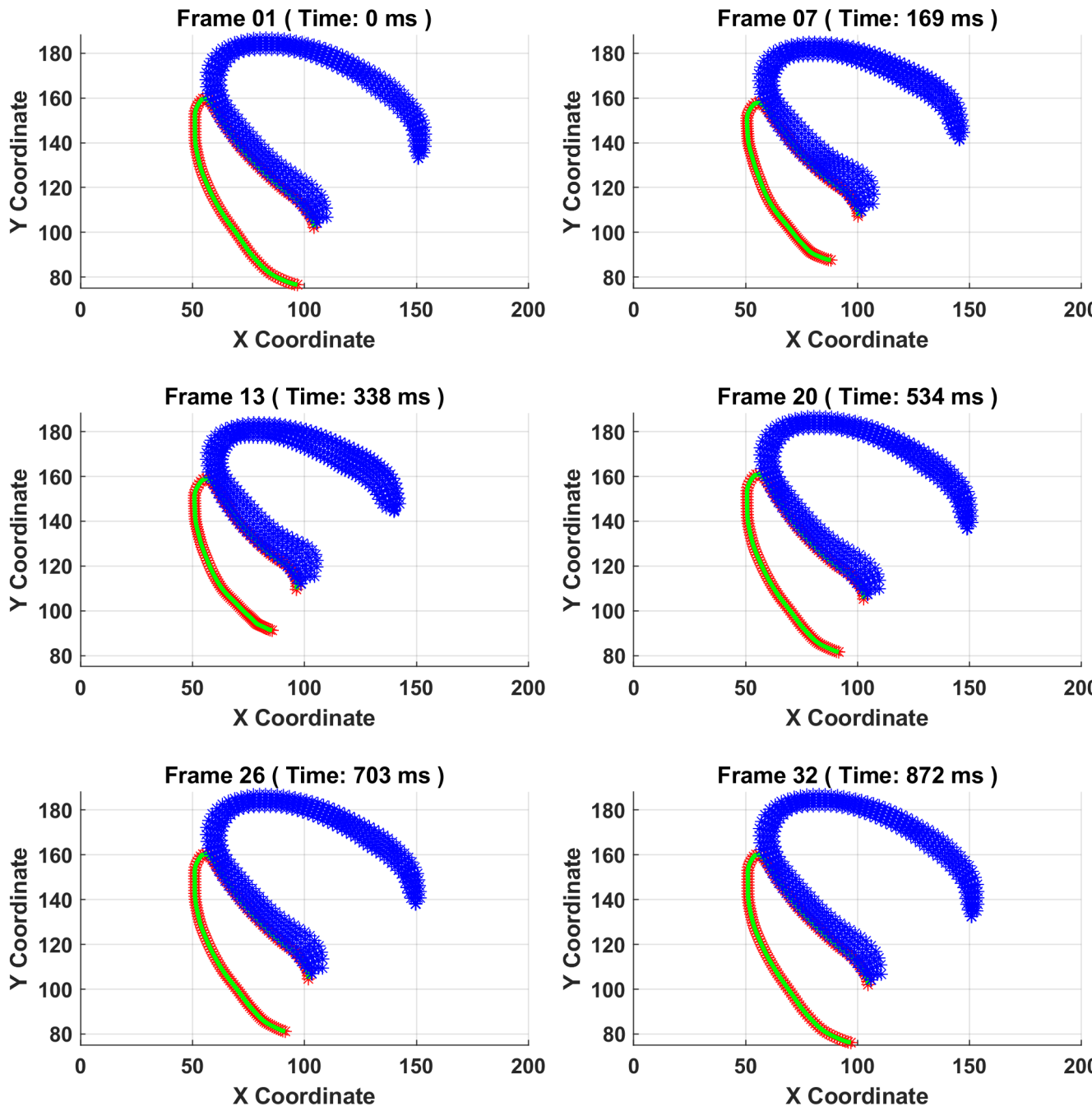

Fig. 9- A long-axis view of ventricular motion during one cardiac cycle, showing the RV in red and the LV in blue.

The following kinematic metrics are expressed to evaluate the LAX slices with the highest kinematic terms, including time-averaged displacement (AD), velocity (AV), and acceleration (AA) over one cardiac cycle.

$$AD = \frac{1}{T}\int_0^T \left|\vec{D}(t)\right| dt \tag{1}$$

$$AV = \frac{1}{T}\int_0^T \left|V(t)\right| dt \tag{2}$$

$$AA = \frac{1}{T}\int_0^T \left|\vec{A}(t)\right| dt \tag{3}$$

In which, $\vec{D}$, $\vec{V}$, and $\vec{A}$ presents time-dependent displacement, velocity, and acceleration, respectively. Moreover, the time duration of one cardiac cycle is represented by $T$. As shown in Fig. 10, for detailed analysis, the RV endocardial surface is segmented into five regions: the basal septum (A), the mid-septum (B), the apex (C), the mid free wall (D), and the basal free wall (E).

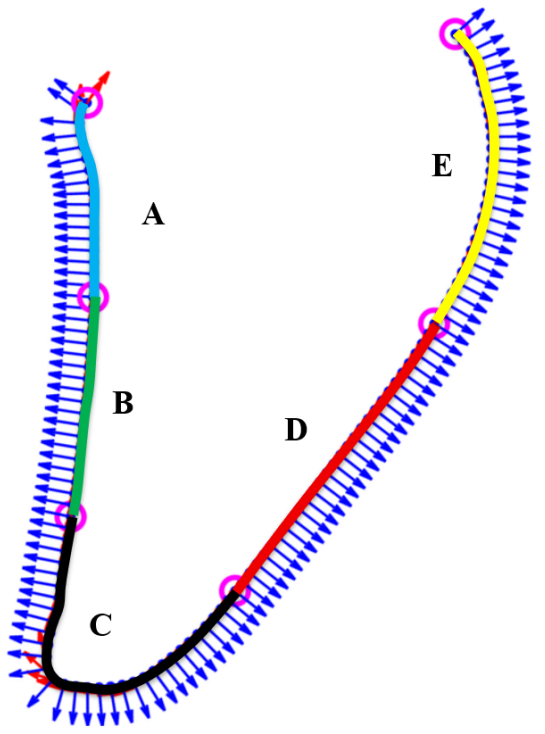

Fig. 10- Right ventricular endocardium segmented into five regions in LAX views: (A) Basal Septum, (B) Mid-Septum, (C) Apex, (D) Mid Free Wall, (E) Basal Free Wall.

The kinematic analysis of each region across different LAX slices is computed using the defined kinematic metrics, as shown in Fig. 11.

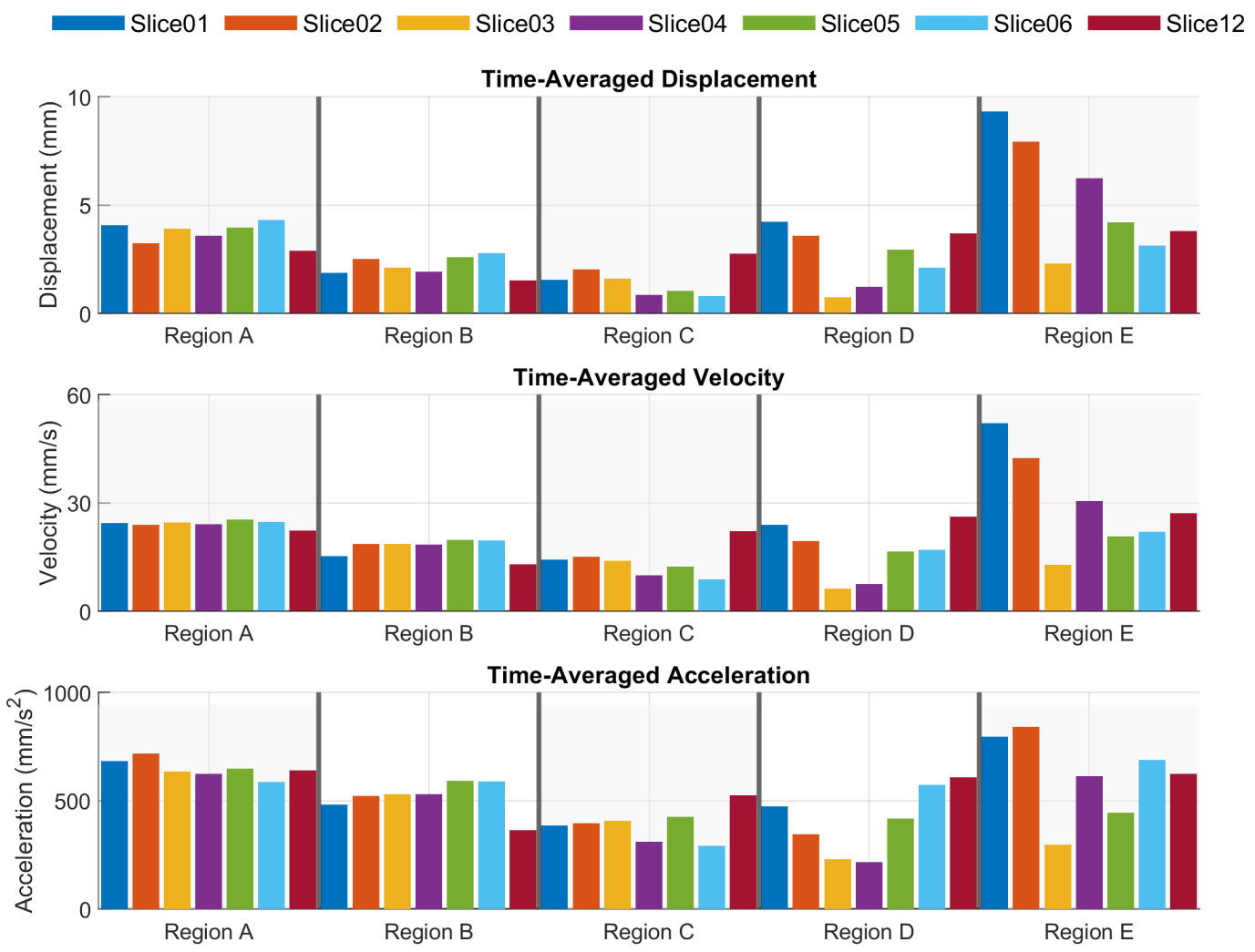

Fig. 11- Time-averaged regional kinematic indices across seven long-axis (LAX) views over one cardiac cycle, comprising mean displacement, velocity, and acceleration for each of the five RV regions (A: Basal Septum, B: Mid-Septum, C: Apex, D: Mid Free Wall, E: Basal Free Wall).

To provide a concise quantitative summary of the regional trends shown in Fig. 11, the average displacement, velocity, and acceleration values were calculated for each RV region and reported in Table I. The results show that Region E exhibits the highest average displacement and velocity, suggesting the greatest overall motion magnitude, whereas Region A shows the highest average acceleration.

TABLE I
AVERAGE DISPLACEMENT, VELOCITY, AND ACCELERATION VALUES ACROSS THE FIVE RV ENDOCARDIAL REGIONS A-E

| Metric | Region A | Region B | Region C | Region D | Region E |
|---|---|---|---|---|---|
| Displacement (mm) | 3.57 | 2.05 | 1.38 | 2.53 | 5.17 |
| Velocity (mm/s) | 23.01 | 16.30 | 12.58 | 15.47 | 28.51 |
| Acceleration (mm/s²) | 636.65 | 501.55 | 377.33 | 394.41 | 600.93 |

To further interpret the slice-dependent trends shown in Fig. 11, the average displacement, velocity, and acceleration were calculated across the five RV regions for each LAX slice and summarized in Table II.

TABLE II
AVERAGE DISPLACEMENT, VELOCITY, ACCELERATION, AND JERK ACROSS THE SELECTED LAX SLICES

| Metric | Slice01 | Slice02 | Slice03 | Slice04 | Slice05 | Slice06 | Slice12 |
|---|---|---|---|---|---|---|---|
| Displacement (mm) | 4.20 | 3.85 | 2.13 | 2.76 | 2.94 | 2.62 | 2.92 |
| Velocity (mm/s) | 25.9 | 23.8 | 15.2 | 18.1 | 18.9 | 18.4 | 22.1 |
| Acceleration (mm/s²) | 565 | 564 | 419 | 458 | 505 | 545 | 551 |

This slice-wise comparison provides a quantitative basis for selecting the view with the greatest motion-related kinetic energy potential. Among the evaluated slices, slice 01 exhibited the highest average displacement and velocity, as well as the highest average acceleration, indicating the strongest overall kinematic response. Therefore, the subsequent analysis in this study focuses on slice 01.

## *III. Regional Motion Validation Against Cine MRI*

The long-axis plane corresponding to the optimal kinematic view (Slice 01) is extracted from the 3D biventricular electromechanical model, as visualized from two perspectives in Fig. 12.

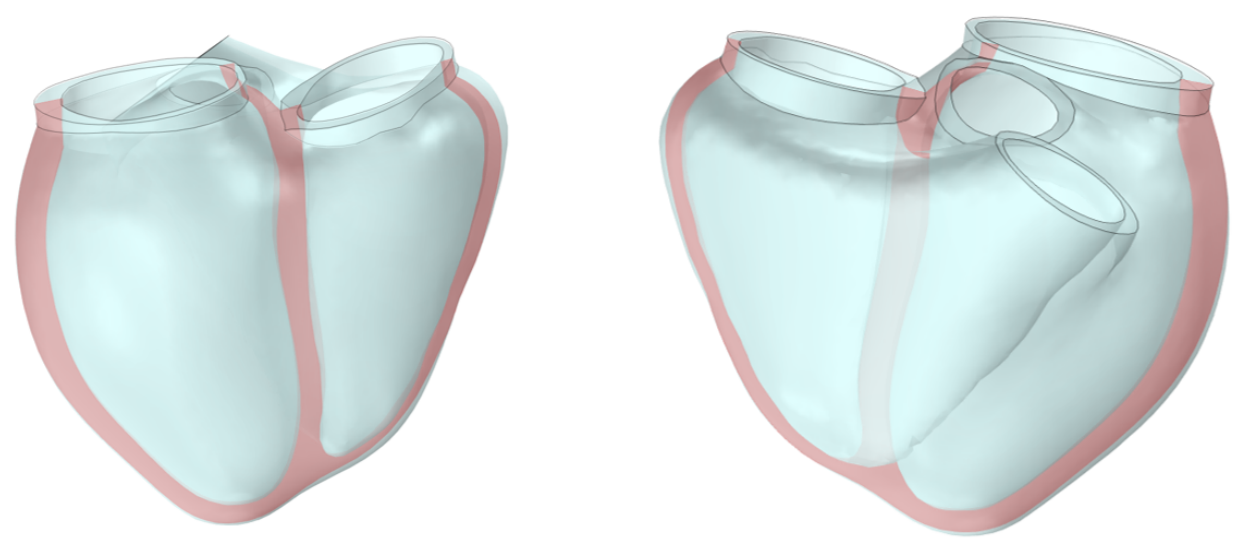

Fig. 12- Two perspective views of a specific long-axis section of the RV, used for a detailed evaluation of its motion throughout the cardiac cycle.

Similarly, the RV surface (in the proposed slice) is subdivided into five specific regions to assess their kinematic energy potential. These regions are shown in Fig. 13, which includes the basal septum (A, red), the mid-septum (B, green), the apex (C, blue), the mid free wall (D, yellow), and the basal free wall (E, purple). This detailed regional segmentation enables the identification of the area exhibiting the most significant motion.

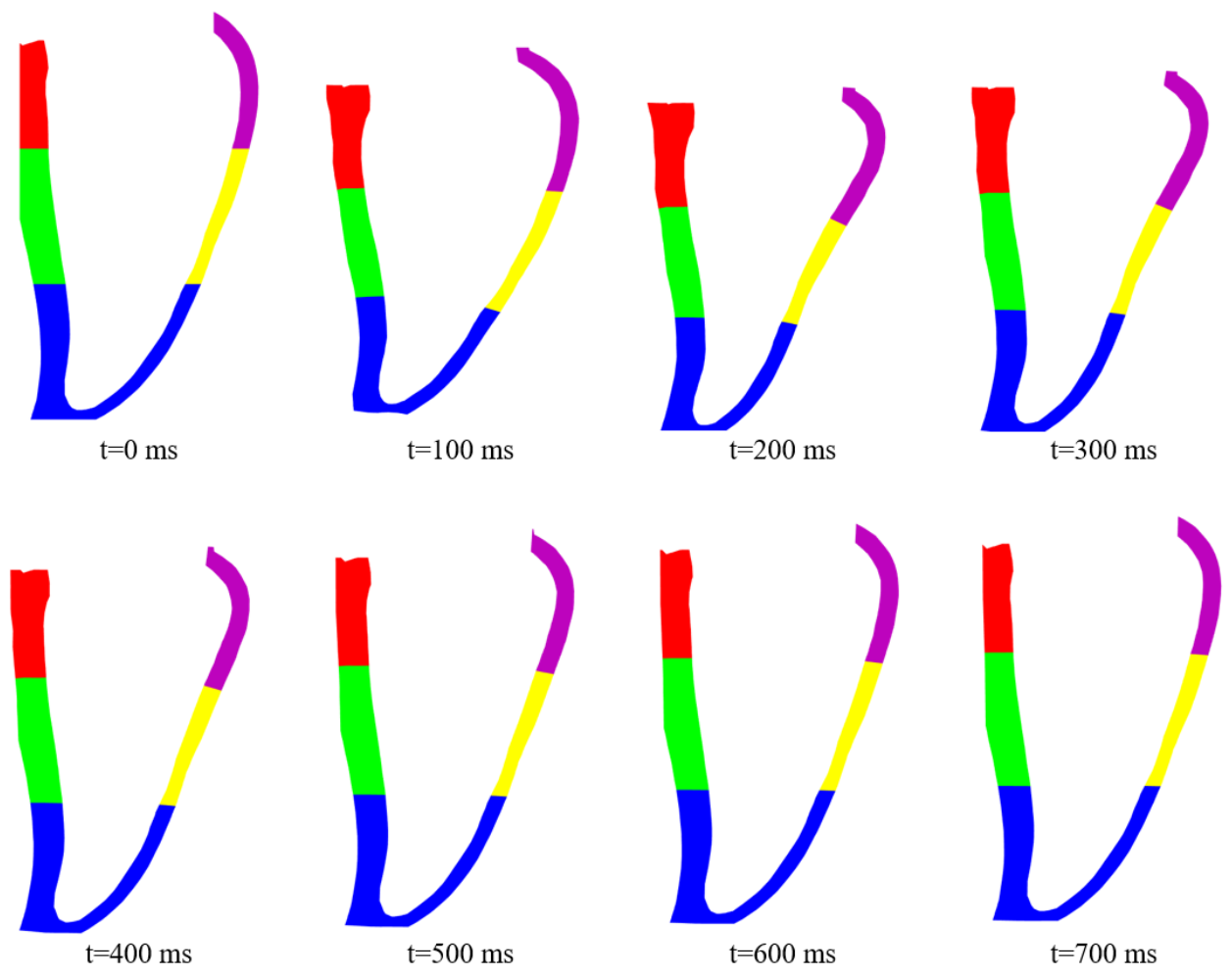

Fig. 13- The 2D representation of the myocardium at the RV during one cardiac cycle; RV is segmented into five regions, including the basal septum (A, red), mid-septum (B, green), apex (C, blue), mid free wall (D, yellow), and basal free wall (E, purple).

For instance, Fig. 14 illustrates the 3D displacement of the basal septum region (A, red) from an arbitrary long-axis view. The presented results reveal that lateral motion (along the Z-axis) is significantly greater than the other directional components.

This observation is consistent with the myocardial fiber architecture shown in Fig. 3, where the right ventricular endocardium exhibits a fiber orientation of approximately +90°. This orientation indicates that the fibers are predominantly aligned along the longitudinal direction of the RV wall, which supports stronger motion and deformation in the basal-to-apical direction. Therefore, the larger displacement observed in the basal and mid free-wall regions can be attributed, in part, to the local endocardial fiber arrangement and its influence on anisotropic myocardial contraction.

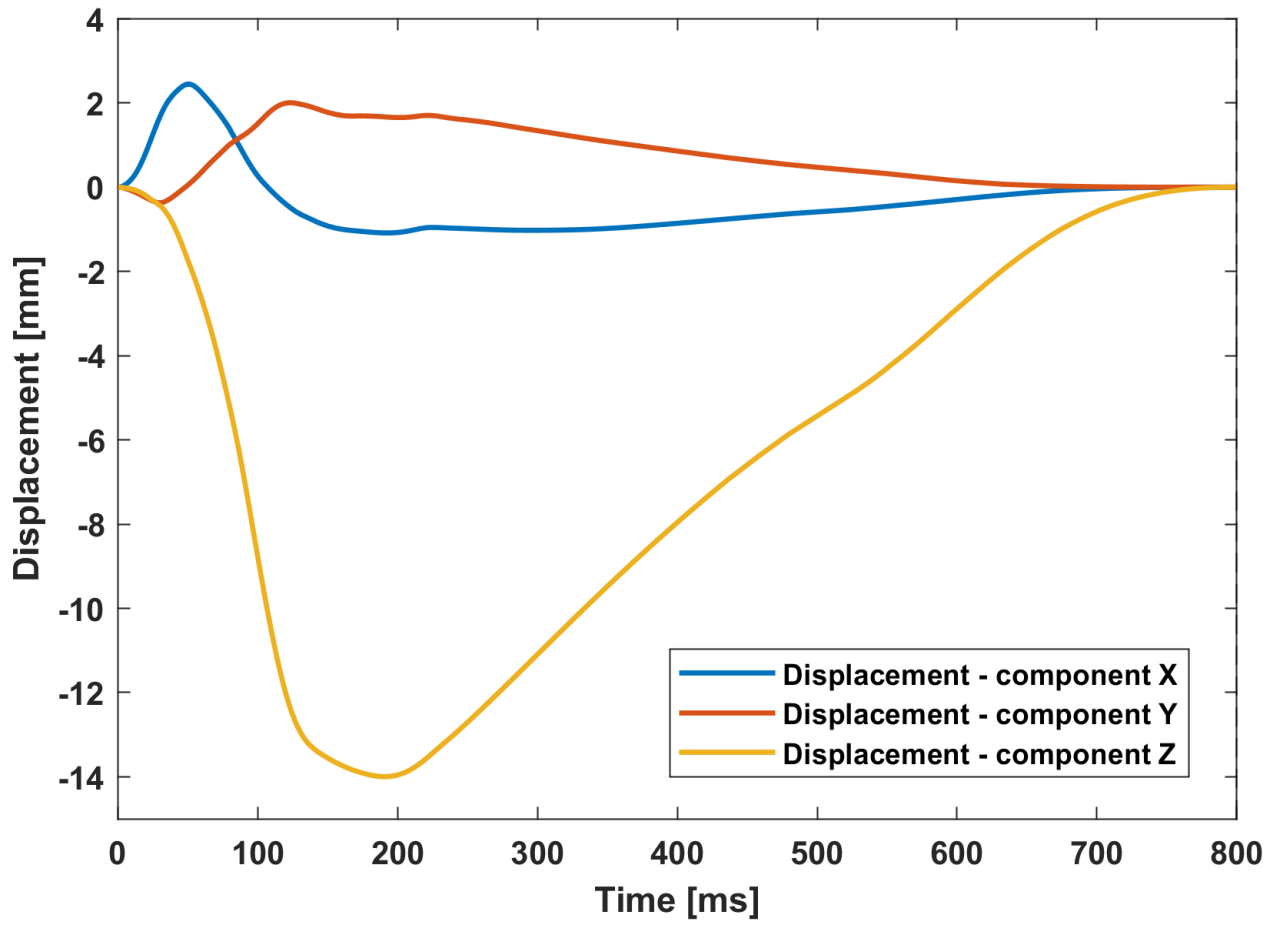

Fig. 14- The 3D displacement of region A, showing dominant lateral motion (Z-axis). This displacement pattern is consistent with the +90° fiber direction in the right ventricular endocardium.

To validate the electromechanical heart model against the Cine-MRI output, the corresponding long-axis plane is isolated from the 3D electromechanical simulation. The simulated right ventricular endocardial wall is tracked using the same five-region segmentation protocol, with regions A, B, C, D, and E to facilitate direct comparison.

Twenty points are selected in each region and tracked to compute spatiotemporal averages of displacement and velocity over one cardiac cycle. Fig. 15(a) and (b) compare the average displacement and velocity of the five regions derived by the developed electromechanical model against the MRI scanning results, respectively. The comparison reveals a consistent regional motion pattern between the electromechanical model and Cine-MRI analysis. The basal free wall (E) exhibits the largest motion magnitude, indicating the highest motion-driven energy harvesting potential. The mid free wall (D) and basal septum (A) exhibit comparable motion levels, with only minor differences, suggesting that both regions provide strong kinematic excitation. In contrast, the apex (C) exhibits the lowest motion level. Therefore, when local cardiac motion is considered alone, the basal free wall remains the most favorable harvesting region, while Regions A and D represent similarly active secondary targets.

However, for an integrated, self-powered leadless pacemaker, the implantation site should not be selected solely based on displacement magnitude. Anatomical safety and electrophysiological suitability should also be considered. The RV free wall is relatively thin and highly deformable, which may increase the risk of local tissue injury or unstable fixation under repeated cyclic loading. In contrast, the basal septum provides a more robust anchoring region and is more compatible with physiological ventricular activation, since native excitation propagates through the septal conduction system before spreading through the ventricular myocardium. Accordingly, although Region E has the highest motion magnitude, Region A represents a potentially favorable implantation region when motion potential, anatomical safety, and electrophysiological suitability are considered together.

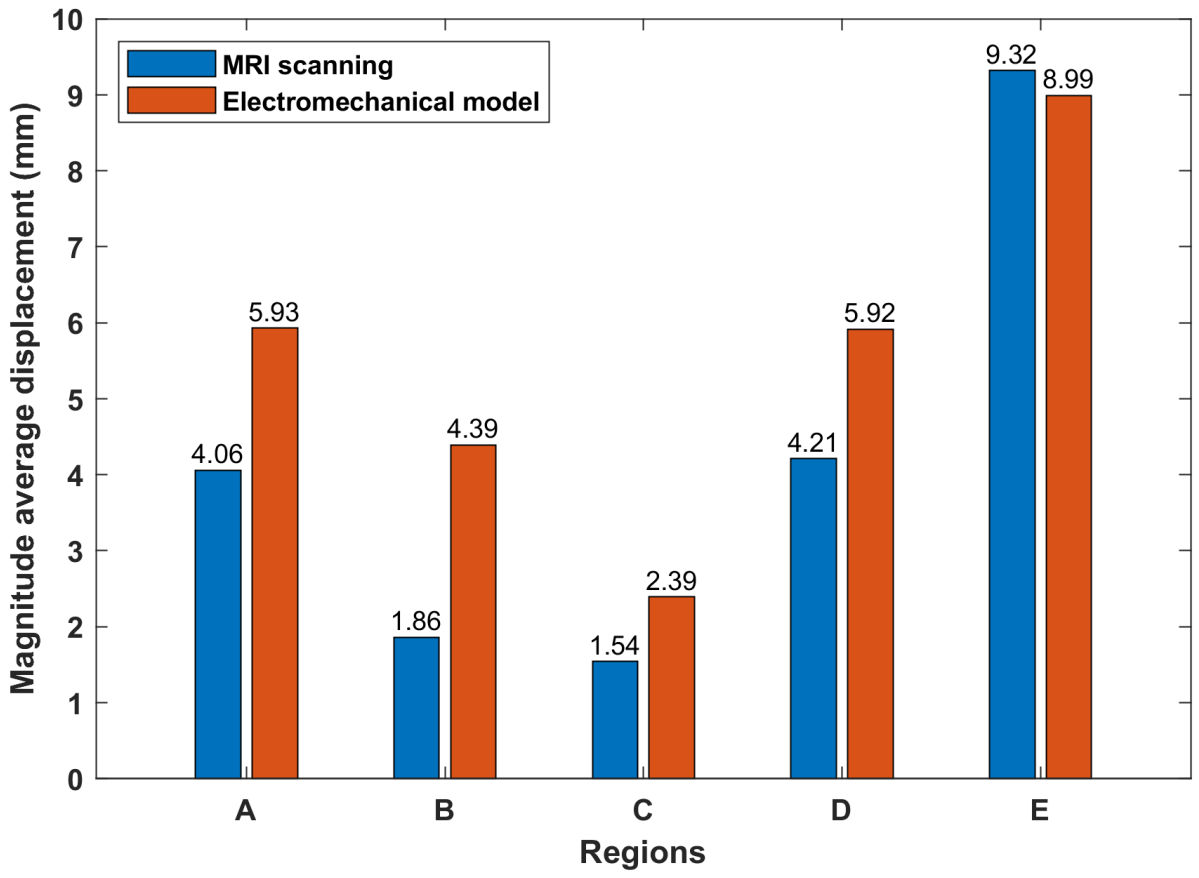

(a)

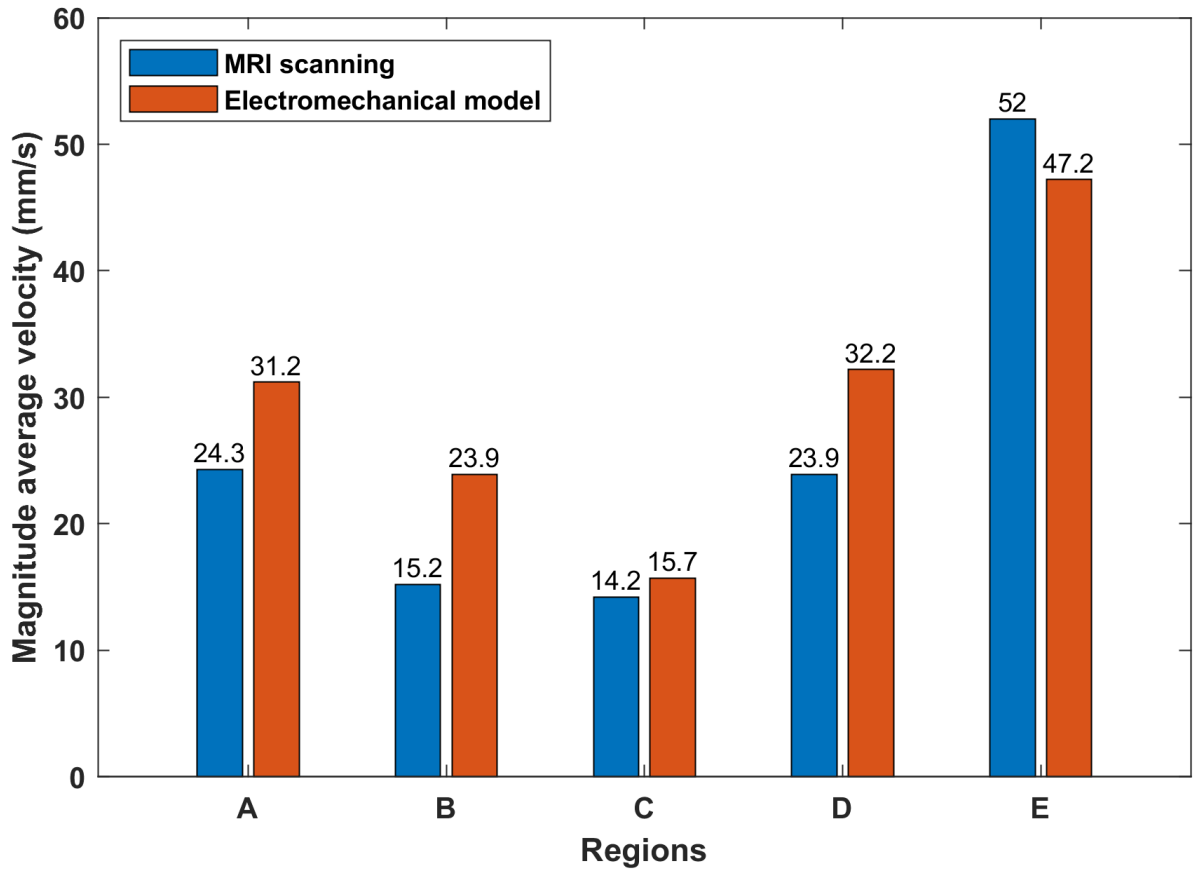

(b)

Fig. 15- The comparison of the time-averaged magnitudes of (a) displacement and (b) velocity predicted by the electromechanical model against MRI scanning results.

Although the electromechanical model and Cine-MRI results show a similar regional motion trend, discrepancies are observed in the absolute displacement and velocity values. These differences are expected because the two approaches represent different subjects and physiological conditions. First, the Cine-MRI data and the anatomical geometry used for the electromechanical model were obtained from different imaging datasets; therefore, subject-specific differences in ventricular shape, wall thickness, and local curvature can affect the measured motion magnitude. Second, the Cine-MRI scan was acquired at a lower heart rate, approximately 66–67 bpm, whereas the electromechanical simulation was performed at 75 bpm, which can alter contraction timing, deformation amplitude, and velocity values. In addition, MRI-derived motion depends on image resolution, slice orientation, segmentation accuracy, and feature-tracking uncertainty, whereas the computational model depends on assumptions about myocardial material properties, fiber orientation, active-stress parameters, hemodynamic loading, and epicardial boundary conditions. Therefore, exact point-by-point agreement in displacement magnitude is not expected. Nevertheless, the model successfully reproduces the main regional ranking observed in MRI.

# IV. Conclusion and outlook

This study developed an MRI-validated biventricular electromechanical heart model for predicting regional endocardial motion. The framework integrates realistic 3D cardiac geometry, rule-based myocardial fiber orientation, electrophysiology, voltage-dependent active stress, two-way fluid–structure interaction, and closed-loop 0D hemodynamics. Validation against Cine-MRI data showed that the model reproduced the main regional motion pattern of the right ventricle, supporting its use as a predictive tool for implantation-site assessment.

The regional analysis showed that the basal free wall (E) exhibited the largest motion magnitude, while the basal septum (A) and mid free wall (D) showed closely comparable kinematic responses. Therefore, if motion amplitude alone is considered, the basal and mid free-wall regions are mechanically favorable for energy harvesting. However, for an integrated self-powered leadless pacemaker, implantation-site selection should also consider anatomical safety and electrophysiological suitability. The RV free wall is relatively thin and highly deformable, whereas the basal septum provides a more robust anchoring site and is more consistent with physiological ventricular activation through the septal conduction pathway.

Accordingly, the results distinguish between the highest-motion harvesting region and the most practical pacemaker implantation site. The basal free wall offers the greatest motion-driven energy harvesting potential, while the basal septum provides a clinically relevant compromise between sufficient kinematic excitation, safer anchoring, and more physiological electrophysiological propagation. Although absolute differences existed between MRI-derived and model-predicted motion values, the agreement in regional ranking supports the validity of the developed model for patient-specific implantation planning.

Future work will focus on incorporating the implant capsule directly into the FSI model to evaluate device–wall interaction, anchoring effects, and changes in local motion and harvestable energy. Further validation across larger healthy and pathological cohorts, together with same-subject anatomical reconstruction, patient-specific physiological calibration, and direct coupling to energy harvester models, will support personalized optimization of self-powered intracardiac devices before clinical implementation.

# Appendix

## A. Electrophysiology parameters: definitions, nominal values, and units

The electrophysiology and active stress parameter values are derived from the literature [31] to reflect human cardiac electrophysiology, which are presented in Table A1.

Table A1- Electrophysiology and active stress parameters: definitions, nominal values, and units

| Parameters | Definition | Value | Unit |
|---|---|---|---|
| **Electrophysiology parameters** | | | |
| $\delta_\phi$ | Additive voltage offset | -80 | mV |
| $\beta_\phi$ | Voltage-scaling factor | 100 | mV |
| $c$ | AP material parameter | 8 | - |
| $a$ | AP threshold parameter | 0.01 | - |
| $\gamma$ | AP recovery parameter | 0.002 | - |
| $\mu_1$ | AP recovery parameter | 0.2 | - |
| $\mu_2$ | AP recovery parameter | 0.3 | - |
| $C_m$ | Membrane capacitance | 1400 | $F/m^2$ |
| $d_{ani}$ | Fiber-aligned anisotropic conductivity coefficient | 0.1 | $mm^2/ms$ |
| $d_{iso}$ | Isotropic conductivity coefficient | 1 | $mm^2/ms$ |
| $\lambda$ | Stretch in the fiber direction | 0.5 | MPa |
| $\phi_s$ | Dimensionless ion channels' resting potential | 0.6 | - |
| $G_s$ | Maximum conductance scaling the stretch-induced current. | 10 | - |
| **Active stress parameters** | | | |
| $\varepsilon_0$ | Lower (baseline) rate constant in the delay function | 0.1 | 1/ms |
| $\varepsilon_1$ | Upper (saturated) rate constant in the delay function | 1 | 1/ms |
| $\vartheta$ | Transition-rate parameter in the switch function | 1 | m/V |
| $\varphi_t$ | Phase-shift (threshold) potential in the delay function | 0 | mV |
| $\nu_f$ | Active stress weight factor - fiber direction | 1.4 | - |
| $\nu_s$ | Active stress weight factor - sheet direction | 0.24 | - |
| $\nu_n$ | Active stress weight factor - sheet-normal direction | 0.9 | - |

## B. Hemodynamics parameters: definitions, nominal values, and units

The lumped-parameter hemodynamic constants for the systemic and pulmonary arterial and venous compartments are presented in Table A2. The parameter's values in the literature [19] were modified to resemble the hemodynamics accurately.

Table A2- The lumped-parameter hemodynamic constants for the systemic and pulmonary arterial/venous compartments

| Parameters | Definition | Value | Unit |
|---|---|---|---|
| **Hemodynamic constant** | | | |
| $\mathbb{R}_{AR}^{SYS}$ | Viscosity resistance of the systemic arteries | 0.8 | mmHg.s.mL$^{-1}$ |
| $\mathbb{R}_{AR}^{PUL}$ | Viscosity resistance of the pulmonary arteries | 0.1625 | mmHg.s.mL$^{-1}$ |
| $\mathbb{R}_{VEN}^{SYS}$ | Viscosity resistance of the systemic veins | 0.26 | mmHg.s.mL$^{-1}$ |
| $\mathbb{R}_{VEN}^{PUL}$ | Viscosity resistance of the pulmonary veins | 0.1625 | mmHg.s.mL$^{-1}$ |
| $\mathbb{C}_{AR}^{SYS}$ | Elastic compliance of systemic arteries | 1.2 | mL.mmHg$^{-1}$ |
| $\mathbb{C}_{AR}^{PUL}$ | Elastic compliance of pulmonary arteries | 10 | mL.mmHg$^{-1}$ |
| $\mathbb{C}_{VEN}^{SYS}$ | Elastic compliance of systemic veins | 60 | mL.mmHg$^{-1}$ |
| $\mathbb{C}_{VEN}^{PUL}$ | Elastic compliance of pulmonary veins | 16 | mL.mmHg$^{-1}$ |
| $\mathbb{L}_{AR}^{SYS}$ | Inertance of blood in systemic arteries | 0.001 | mmHg.s$^2$.mL$^{-1}$ |
| $\mathbb{L}_{AR}^{PUL}$ | Inertance of pulmonary arteries. | 0.002 | mmHg.s$^2$.mL$^{-1}$ |
| $\mathbb{L}_{VEN}^{SYS}$ | Inertance of systemic veins | 0.005 | mmHg.s$^2$.mL$^{-1}$ |
| $\mathbb{L}_{VEN}^{PUL}$ | Inertance of pulmonary veins | 0.002 | mmHg.s$^2$.mL$^{-1}$ |
| $\rho_{fluid}$ | Blood density | 1060 | Kg/m$^3$ |
| $\nu_{fluid}$ | Dynamic viscosity of blood | 0.004 | Pa.s |
| **Valves parameters** | | | |
| $R_O$ | Valve resistance during the open phase | 0.0075 | mmHg.s.mL$^{-1}$ |
| $R_C$ | Valve resistance during the closed phase | 75000 | mmHg.s.mL$^{-1}$ |
| $\Delta p_{tran}$ | Transvalvular pressure difference | 0.45 | mmHg |

## C. Coefficients of mechanical boundary conditions

The following table lists the coefficients for the mechanical boundary conditions applied at epicardial and basal boundaries.

Table A3- The coefficients of mechanical boundary conditions.

| Parameters | Value | Unit |
|---|---|---|
| $K_t^{\Gamma^{epi}}$ | 2e5 | N/(m.m$^2$) |
| $K_n^{\Gamma^{epi}}$ | 2e4 | N/(m.m$^2$) |
| $K_l^{\Gamma^{epi}}$ | 2e4 | N/(m.m$^2$) |
| $C_t^{\Gamma^{epi}}$ | 2e4 | N.s/(m.m$^2$) |
| $C_n^{\Gamma^{epi}}$ | 2e3 | N.s/(m.m$^2$) |
| $C_l^{\Gamma^{epi}}$ | 2e3 | N.s/(m.m$^2$) |
| $k_t^{\Gamma^{base}}, k_n^{\Gamma^{base}}, k_l^{\Gamma^{base}}$ | 0 | N/(m.m$^2$) |
| $c_t^{\Gamma^{base}}, c_n^{\Gamma^{base}}, c_l^{\Gamma^{base}}$ | 2e4/3 | N.s/(m.m$^2$) |